\documentclass{SCGE}
\usepackage[colorlinks,linkcolor=blue,citecolor=blue,urlcolor=blue]{hyperref}
\usepackage[utf8]{inputenc}
\usepackage{rotating}
\usepackage{footnote}
\usepackage{times}
\usepackage{newtxtext,newtxmath}
\usepackage{graphicx}	
\usepackage{amsmath}	
\usepackage{amssymb}	
\usepackage{amsthm}
\usepackage{mathtools}
\usepackage[normalem]{ulem}
\usepackage{gensymb}

\usepackage{CJK}
\usepackage{multirow, array}

\let\citedash\relax
\makeatletter \providecommand{\citedash}{\hbox{-}\penalty\@m}
\makeatother

\begin{document}

\begin{picture}(0,0){\rm
\put(0,-20){\makebox[160truemm][l]{\bf {\sanhao\raisebox{2pt}{.}}
Article  {\sanhao\raisebox{1.5pt}{.}}}}}
\put(0,-34){\jiuwuhao {\textcolor[rgb]{0.5,0.5,0.5}{\sf 
}}}
\end{picture}



\Year{2021} %
\Month{XXXX} %
\Vol{xx} %
\No{x} %
\BeginPage{1} %
\AuthorMark{{\rm C. Yang et al.} }  
\DOI{} 
\ArtNo{000000}

\title{Investigating the co-evolution of massive black holes in dual active galactic nuclei and their host galaxies via galaxy merger simulations}

\author[1, 2, 3*]{Chao Yang}{}
\author[1, 3]{Junqiang Ge}{}
\author[1, 2, 3*]{Youjun Lu}{}
\footnote{*Corresponding author (Youjun Lu, email: luyj@nao.cas.cn; \\
Chao Yang, email: chyang@nao.cas.cn)}

\address[{\rm1}]{National Astronomical Observatories, Chinese Academy of Sciences, Beijing 100012, China;}
\address[{\rm2}]{School of Astronomy and Space Sciences, University of Chinese Academy of Sciences, Beijing 100049, China;}
\address[{\rm3}]{CAS Key laboratory for computational Astrophysics, National Astronomical Observatories, Chinese Academy of Sciences, Beijing, 100012, China}


\maketitle \vspace{-3.5mm}{\footnotesize\begin{center} Received Month date, Year; accepted Month date, Year
\end{center}}\vspace*{-5mm}

\begin{center}
\rule{16.5cm}{0.4pt}
\parbox{16.5cm}
{\begin{abstract} 
Major galaxy mergers can trigger nuclear activities and are responsible for high-luminosity quasi-stellar objects /active galactic nuclei (QSOs/AGNs). In certain circumstances, such mergers may cause dual active galactic nuclei (dAGN) phenomenon. This study investigates dAGN triggering and evolution of massive black holes (MBHs) during the merging processes using hydrodynamic code GADGET-2 to simulate several gas-rich major mergers at redshift $z=2$ and $3$, respectively. Results reveal that gas-rich major mergers can trigger significant nuclear activities after the second and third pericentric passages and the formation of dAGN with significant time duration ($\sim 10 - 390$\,Myr). During the merging processes, galactic bulge evolves with time because of the rapid star formation in each (or both) galactic centers and initial mixing of stars in galactic disks due to violent relaxation. MBHs grow substantially due to accretion and finally merge into a bigger black hole. The growth of galactic bulges and corresponding increases of its velocity dispersions predate the growth of MBHs in the dAGN stages. The MBHs in these stages deviate below the relation between MBH mass and bulge mass (or velocity dispersion), and they revert to the relation after the final mergers due to the significant accretion that occurs mostly at a separation less than a few kpc. Then, the two MBHs merge with each other.

\end{abstract}}
\end{center}\vspace*{-0.6cm}
\begin{center}
\parbox{16.5cm}
{\bf\jiuhao Galaxies, quasars, black hole, numerical simulations}
\end{center}

\begin{center}
{}
\end{center}

\textwidth=178truemm \textheight=236truemm%

\wuhao\vspace*{1.5mm}

\begin{multicols}{2}

\renewcommand{\baselinestretch}{1.08} \baselineskip 12.2pt\parindent=10.8pt

\renewcommand{\thefootnote}

\section{Introduction}
\label{sec:intro}

Nuclear activities are triggered by the infall of gaseous material to the vicinity of massive black holes (MBHs) in galactic centers \cite{1964ApJ...140..796S, 1969Natur.223..690L, 1999agnc.book.....K}. The energy and momentum output from those active galactic nucleus (AGNs) may regulate the evolution and growth of their host galaxies by heating up and/or removing cold gas, thus suppressing star formation in those galaxies \cite{1998A&A...331L...1S, 2003ApJ...596L..27K, 2012ARA&A..50..455F, 2014ARA&A..52..589H, 2015ARA&A..53..115K}. 
The close relations between the masses of MBHs and the properties of their host galaxies \cite{1998AJ....115.2285M, 2000ApJ...539L...9F, 2000ApJ...539L..13G, 2002ApJ...574..740T, 2013ARA&A..51..511K, 2016ASSL..418..263G} are suggested to be the natural results of feedback processes in the scenario of co-evolution of galaxies and their central MBHs.

Galaxy mergers play a central role in the hierarchical galaxy formation and MBH growth under the $\Lambda$CDM cosmogony \cite{2005Natur.435..629S}. Numerical simulations have suggested that major mergers can cause a gas to sink to galactic centers, trigger nuclear activity, and cause phenomena such as QSOs and AGNs \cite{1989Natur.340..687H, 2011MNRAS.418.2043E, 2017ApJ...838..129B, 2018MNRAS.476.2308W}. However, observational evidence for the connection between mergers of galaxies and QSOs/AGNs is still inconclusive. Some observations on the host galaxies of QSOs/AGNs have revealed that some host galaxies have considerably disturbed morphology probably because of recent major mergers \cite{2011MNRAS.418.2043E, 2012ApJ...758L..39T, 2014A&A...569A..37M, 2014MNRAS.441.1297S, 2015ApJ...804...34H, 2018ApJ...853...63D, 2018PASJ...70S..37G}. Other studies reported that the host galaxies of many low redshift low luminosity AGNs have no merger signatures but have disky structures, which may suggest no connection with major mergers. These AGNs are proposed to be triggered by secular processes or minor mergers rather than major mergers \cite{2011ApJ...726...57C, 2012ApJ...744..148K, 2017MNRAS.470..755H, 2017MNRAS.465.2895L, 2019MNRAS.483.2441V}. These observational results prevent a clear understanding of the triggering mechanism(s) of nuclear activities and the co-evolution of MBHs and galaxies.

During the merger process of two galaxies, one of the prominent phenomena might be the ignition of both MBHs, with which the system appears as a dual AGN (dAGN; \cite{2003ApJ...582L..15K, 2011ApJ...735L..42K, 2012ApJ...746L..22K, 2012ApJ...748L...7V}).
dAGNs are idea targets to study the relation between major mergers and nuclear activities as well as the co-evolution of MBHs and host galaxies as they are caught in action. 
Approximately 30 dAGNs (or dAGN candidates) have been obtained by adopting different methods and signatures, which result in about $30$ dAGNs (or dAGN candidates). These methods include (1) identifying two spatially separate line cores of those systems with double-peaked narrow emission lines \cite{2004ApJ...604L..33Z, 2009ApJ...705L..76W, 2009ApJ...705L..20X, 2009ApJ...698..956C, 2010ApJ...708..427L, 2010ApJ...715L..30L, 2011ApJ...733..103F, 2012ApJ...745...67F, 2012ApJS..201...31G, 2013MNRAS.429.2594B, 2016MNRAS.457.3878Z, 2018ApJ...867...66C, 2019MNRAS.482.1889W, 2019arXiv190406716W}, (2) identifying two spatially separate X-ray cores in a merging galaxy \cite{2003ApJ...582L..15K, 2011ApJ...735L..42K, 2011ApJ...737L..19C, 2012ApJ...746L..22K}, and (3) identifying two separate radio cores in a merging galaxy  \cite{2011ApJ...740L..44F, 2012MNRAS.425.1185F, 2015ApJ...813..103M}, etc. However, obtaining a large sample of dAGNs is difficult because of various problems. For example, the method utilizing double-peaked narrow emission lines is inefficient in identifying dAGNs although many AGNs have such emission lines formed because of the majority of those AGNs with such a feature are caused by outflow or rotating disk \cite{2005ApJ...627..721G, 2011ApJ...735...48S}. Current X-ray and radio surveys obtained only a limited number of low-redshift AGNs having sufficiently high spatial resolution \cite{2017ApJ...848..126S}. Despite of these difficulties, it is anticipated that the number of dAGNs is anticipated to considerably increase and provide meaningful statistics about their properties by using future powerful telescopes and conducting large area surveys \cite{2018ApJ...854..169L, 2018ApJ...862...29L}.

Previous theoretical studies have been performed in the past to predict the properties, duty cycle and frequency of dAGNs, by either numerically simulating \cite{2012ApJ...748L...7V, 2013MNRAS.429.2594B} or semi-analytically modeling \cite{2011ApJ...738...92Y} merging galaxies. Both hydrodynamic simulations and semi-analytical modeling are essential for our understanding of dAGNs. The former one provides detailed information on individual mergers, such as how and when the nuclear activity is triggered, how long does such activity is maintained, and how do MBHs co-evolve with host galaxies during the merging processes. The latter one may provide statistical estimates on the frequency of dAGNs and distributions of various properties of dAGNs, as a supplement to the former one, given that high-resolution hydrodynamic simulations can be performed for only a limited number of mergers because they are too expensive. A comparison of theoretical studies with observational results on dAGNs is expected to provide important information and place strong constraints on the triggering mechanism of nuclear activity and the co-evolution of MBHs and galaxies.

This study investigates the ignition of dAGNs and the co-evolution of MBHs in dAGN with their host galaxies by using high resolution hydrodynamic simulations of several merging galaxies. This paper is organized as follows: In Section~\ref{sec:methods}, we briefly introduce the initial settings for our galaxy merger simulations and provide a summary of the methods that are adopted to analyze the simulation data. In Section~\ref{sec:result}, we present the simulation results on the ignition of MBH accretion/dAGN and star formation during the galaxy merging process, and illustrate the evolution of the relationship between MBH mass and host galaxy properties, e.g., the bulge mass or bulge velocity dispersion. In Section~\ref{sec:summary}, we present the conclusions and discussions.

\section{Numerical simulations}
\label{sec:methods}

\subsection{Initial setup}

We adopt the smoothed particle hydrodynamics (SPH) code GADGET-2 \cite{2005MNRAS.364.1105S} to simulate the galaxy merging processes.
We also implement a number of physical processes into the GADGET-2 code to study the evolution/growth of bulges and MBHs, and the triggering of nuclear activities, i.e., star formation, MBH accretion, star formation, supernova and AGN feedback. 

The star formation process and supernova feedback are implemented based on the hybrid model described in \cite{2003MNRAS.339..289S}. In this model, a gas particle contains a hot component and a cold component. The gas density is defined as $\rho_{\mathrm{gas}} = \rho_{\mathrm{h}} + \rho_{\mathrm{c}}$, where $\rho_{\mathrm{h}}$ and $\rho_{\mathrm{c}}$ are the density of hot and cold gas, respectively. The star formation rate at a characteristic timescale $t_{*}$ is
\begin{equation}
\frac{\mathrm{d}\rho_{*}}{\mathrm{d}t} = (1 - \beta)\frac{\rho_{\mathrm{c}}}{t_{*}},
\end{equation}
where $\beta$ is the mass fraction of the stars that exploded as supernovae instantly. The star formation is regulated by a density threshold $\rho_{\mathrm{th}}$ to match the observation \cite{2003MNRAS.339..289S, 2004MNRAS.348..435N, 2014ApJ...780..145T}. 

On the basis of the Bondi$-$Hoyle$-$Lyttleton parameterization \cite{1939PCPS...35..405H, 1944MNRAS.104..273B, 1952MNRAS.112..195B}, the MBH accretion rate is defined as:
\begin{equation}
\dot{M} = \frac{4 \pi \alpha G^{2} M_{\bullet}^{2} \rho_{\mathrm{gas}}}{(c_{\mathrm{s}}^{2} + v^{2})^{3/2}},
\label{eq:Mgrow}
\end{equation}
where $M_{\bullet}$ is the BH mass, $c_{\mathrm{s}}$ is the sound speed of the gas, $v$ is the velocity of the BH relative to the gas, and $\alpha$ is a dimensionless parameter. As in \cite{2009MNRAS.398...53B} and \cite{2009ApJ...690..802J}, here we similarly set $\alpha = 8$ to ensure a reasonable BH accretion rate, which enables us to avoid a relative high accretion rate at the beginning of the simulation and to check whether the galaxy merger can trigger a strong BH activity. Different choices of $\alpha$, typically in the range from 1 to 100, are discussed in the literature\cite{2014MNRAS.442.2751T, 2005MNRAS.361..776S, 2015MNRAS.446..521S}. However, it does not introduce significant effect on the MBH accretion and growth. Choosing a higher $\alpha$ mainly results in two kinds of effects: 1) more significantly suppressing the star formation by influencing the gas cooling in the central region of the host galaxy (in $\alpha = 100$ case, the SFR can be suppressed to three times lower than the $\alpha = 1$ case)\cite{2014MNRAS.442.2751T}, and 2) helping the MBH enter the Eddington-limited regime quickly in the case of low spatial resolution \cite{2005MNRAS.361..776S, 2015MNRAS.446..521S} (if the BH is self-regulated, then the energy released as feedback that is independent of $\alpha$). Therefore, the overall MBH growth over cosmic time is independent of $\alpha$ \cite{2014MNRAS.442.2751T}. Meanwhile, we set the Eddington accretion rate $\dot{M}_{\mathrm{Edd}}$ as the upper limit of the accretion rate to guarantee that no super-Eddington accretion appears. The energy injection rate of the MBH accretion process to the surrounding gas, i.e., the AGN feedback rate, is given by
\begin{equation}
  \dot{E}_{\rm feed} = \epsilon_{\rm f} L_{\rm bol} = \epsilon_{\rm f} \epsilon_{\rm r} \dot{M} c^2,
\end{equation}
where $\epsilon_{\mathrm{f}}$ is the fraction of the bolometric luminosity $L_{\mathrm{bol}}$ that is deposited into the surrounding gas, $\epsilon_{\mathrm{r}}$ is the mass-to-energy conversion efficiency, and $c$ is the speed of light. In this work, we set $\epsilon_{\mathrm{r}} = 0.1$ and $\epsilon_{\mathrm{f}} = 0.05$, which are typical values for the studying of AGN feedback \cite{2005Natur.433..604D, 2017SCPMA..60j9511Z}. As to the black hole accretion procedure, we can neither resolve the Bondi radius ($\lesssim 1$ pc for $M_{\bullet}\lesssim 10^7 M_{\odot}$) nor describe the accretion flow in details due to the limited spatial resolution ($\ge 10$ pc) of the current simulation. Therefore, in our simulation we set the black accretion process by following the coarse-graining procedure as usually assumed in the GADGET-2 simulation \cite{2005ApJ...630..705H, 2005Natur.433..604D, 2017MNRAS.468.3395M, 2019MNRAS.483.4640W}, in which the central MBH has both the increased mass and momentum as a consequence of the gas accretion and a fraction of radiation energy can affect the surrounding gas as feedback. The numerical implementation of MBH accretion and AGN feedback are taken from the procedures detailed in \cite{2005MNRAS.361..776S}.

To study the galaxy-MBH co-evolution scenario of spiral galaxies, we run three spiral-spiral galaxy mergers with mass ratio $1:2$, given that gas rich major mergers are expected to trigger nuclear activities. We note here that the major mergers were believed to result in the destruction of the disky structure of a galaxy and thus an elliptical galaxy. However, recent studies show that a disk galaxy can still be produced by a major merger if the interacting galaxies are gas rich and have sufficiently large orbital angular momentum \cite{2005ApJ...622L...9S, 2017MNRAS.470.3946S}. 
With the primary and secondary galaxies built up, the three spiral-spiral galaxy mergers are designed as follows: one simulation starts at $z = 2$ and another at $z = 3$ with a parabolic orbit of two progenitor galaxies, and the merger is set to be co-planar (with an inclination angle of $i=0^\circ$) and prograde. The inclination angle $i$  defined here is the angle between the disk normals of the two progenitor galaxies, and $i=0^{\circ}$ means the stellar disks of the two galaxies are on the same plane. To study the effect of galaxy orientation on the bulge formation, the third simulation is set to start at $z=3$ but with $i=45^{\circ}$.  With these setups, we can check both the difference between the mergers started at different redshifts and mergers that started at the same redshift but with different inclined angles.

\begin{table*}
\caption{Physical parameters for the progenitor galaxies of mergers
}
\begin{center}
\begin{tabular}{cccccc}
\hline
\multicolumn{1}{c}{Symbol} & \multicolumn{2}{c}{$z = 2$} & \multicolumn{2}{c}{$z = 3$} & Description\\
\cmidrule(r){2-3}  \cmidrule(r){4-5}
& Primary & Secondary & Primary & Secondary & \\
\hline
$M_{\mathrm{vir}}$ & $2.24 \times 10^{11} M_{\odot}$ &  $1.12 \times 10^{11} M_{\odot}$ &  $2.24 \times 10^{11} M_{\odot}$ &  $1.12 \times 10^{11} M_{\odot}$ & Virial mass of the primary galaxy\\
$m_{\mathrm{d}}$ & 0.04$M_{\mathrm{vir}}$ &  0.04$M_{\mathrm{vir}}$ &  0.04$M_{\mathrm{vir}}$ &  0.04$M_{\mathrm{vir}}$ & Disk mass\\
$f_{\mathrm{gas}}$ & 0.3 &  0.3 &  0.3 &  0.3 & Gas fraction in the disk\\
$m_{\mathrm{b}}$ & 0.008$M_{\mathrm{vir}}$ &  0.008$M_{\mathrm{vir}}$ &  0.008$M_{\mathrm{vir}}$ &  0.008$M_{\mathrm{vir}}$ & Bulge mass \\
$M_{\mathrm{BH}}$ & $3 \times 10^{6} M_{\odot}$ &  $1.5 \times 10^{6} M_{\odot}$ & $3 \times 10^{6} M_{\odot}$ & $1.5 \times 10^{6} M_{\odot}$ &BH mass\\

$j_{\mathrm{d}}$ & 0.04$j_{\mathrm{halo}}$ &  0.04$j_{\mathrm{halo}}$ &  0.04$j_{\mathrm{halo}}$ &  0.04$j_{\mathrm{halo}}$ & Spin of the disk\\
$R_{200}$ & 61.0 $\mathrm{kpc}$ & 48.2 $\mathrm{kpc}$ & 46.4 $\mathrm{kpc}$ & 37.0 $\mathrm{kpc}$ & Halo Virial radius\\
$R_{\mathrm{H}}$ & 10.53 $\mathrm{kpc}$ & 8.32 $\mathrm{kpc}$ & 8.66 $\mathrm{kpc}$ & 6.84 $\mathrm{kpc}$ & Scale radius of Hernquist profile\\
$R_{\mathrm{S}}$ & 6.1 $\mathrm{kpc}$ & 4.8 $\mathrm{kpc}$ & 5.17 $\mathrm{kpc}$ & 4.08 $\mathrm{kpc}$ & Halo scale radius\\
$H$ & 1.17 $\mathrm{kpc}$ & 0.92 $\mathrm{kpc}$ & 0.94 $\mathrm{kpc}$ & 0.74 $\mathrm{kpc}$ & Disk scale length\\
$Z_{0}$ & 0.12 $\mathrm{kpc}$ & 0.09 $\mathrm{kpc}$ & 0.10 $\mathrm{kpc}$ & 0.07 $\mathrm{kpc}$ & Disk thickness\\
$A$ & 0.23 $\mathrm{kpc}$ & 0.18 $\mathrm{kpc}$ & 0.19 $\mathrm{kpc}$ & 0.15 $\mathrm{kpc}$ & Bulge scale radius\\
\hline
\end{tabular}
\end{center}
\label{tab:ini_galaxy}

\end{table*}

The virial mass of the primary galaxy is $M_{\mathrm{vir}} = 2.24 \times 10^{11}M_{\odot}$. For each isolated galaxy, we assume the disk as exponential, and both the bulge and dark matter halo follow the Hernquist profile \cite{1990ApJ...356..359H}. In the galactic disk, we set the gas fraction $f_{\mathrm{gas}} = 0.3$; the rest are stars. The mass fraction of the disk to the galaxy virial mass is $m_{\mathrm{d}} = 0.04$ and the spin of the disk $j_{\mathrm{d}}$ is taken as $4\%$ of the dark matter halo spin $j_{\mathrm{halo}}$, i.e., $j_{\mathrm{d}} = 0.04 j_{\mathrm{halo}}$, where $j_{\mathrm{halo}}$ is calculated by following the definition in \cite{1998MNRAS.295..319M}. A spherical bulge is also placed around the galactic center, the mass fraction of the bulge is $m_{\mathrm{b}} = 0.008$. At each galactic center, we place a sink particle as the MBH, which may accrete the surrounding gas particle and interact with its host galaxy through the AGN feedback process. The masses of MBHs of the primary and secondary galaxies are $3 \times 10^{6} M_{\odot}$ and $1.5 \times 10^{6} M_{\odot}$, respectively. The mass of each MBH is set by the $M_{\bullet}-M_{\mathrm{b}}$ relation \cite{2003ApJ...589L..21M}. The initial parameter setup of our simulation is
summarized in Table \ref{tab:ini_galaxy}, and the corresponding mass and spatial resolutions of those simulations are listed in Table \ref{tab:ini_setup}.

The initial separation of the two progenitor galaxies in all the three simulations is set as the sum of the virial radii of the two galaxies. The pericentric distance is $20\%$ of the virial radius of the primary galaxy. To present sufficient details of the merging processes, we output the snapshot of the simulation in each 4 Myrs, which allows for statistical analyses of the BH accretion and star formation processes.

\subsection{Dual AGN}

Nuclear activities can be triggered during the galaxy merging process. If both MBHs are active at the same time and separated from each other on the scale of kpc, the system may appear as a dAGN \cite{2003ApJ...582L..15K, 2012ApJ...748L...7V, 2013MNRAS.429.2594B}. To match the current observational capability of most telescopes and make our results comparable with those from observations \cite{2012ApJ...744....2A, 2019ApJ...870...31I} or other simulations \cite{2012ApJ...748L...7V, 2017MNRAS.469.4437C}, 
through out this paper, we choose two thresholds in the bolometric luminosity (either $L_{\mathrm{bol}} = 10^{44}\mathrm{erg\ s}^{-1}$ or $L_{\mathrm{bol}} = 10^{43}\mathrm{erg\ s}^{-1}$), two thresholds in the Eddington ratio ($f_{\mathrm{Ed}} = 0.05$, $f_{\mathrm{Ed}} = 0.1$), and two separation thresholds ($r>0.1$ kpc, $r>1$ kpc) to match different spatial resolutions of current telescopes. A merging system is classified as a dAGN if the two accreting MBHs that are hosted in it can have luminosities or Eddington ratios above these thresholds in a proper separation. These thresholds are chosen for the convenience of comparing our simulation results with observations \cite{2012ApJ...744....2A, 2019ApJ...870...31I} or other simulations \cite{2012ApJ...748L...7V, 2017MNRAS.469.4437C}.

\subsection{Bulge mass and velocity dispersion}

In order to study the evolution of the relationships between MBH mass
and host galaxy properties during the merger process, especially at the period of dAGNs, we decompose the bulge component associated with each of the two MBHs at different stages of the merger. 2D image fitting is inaccurate for performing the decomposition because the galactic disk is highly perturbed during the merger. Therefore, we decompose the galaxy profile by using long slits in different directions to obtain the effective radius of the bulge components associated with each MBH. We take a long slit across the two MBHs and other slits perpendicular to the first slit and across these two MBHs, respectively. We exclude the density information in between these two MBHs because it is perturbed and the most and difficult to separate into two components associated with each MBH especially at the late merger stage. We obtain three density profiles for each progenitor around each MBH and stack them together to finally obtain an average density. The width of each slit width is set to $0.2$\,kpc, and the radial bins are also set to $0.2$\,kpc. Each profile is  truncated at a radius where the particle number in that radius bin is less than $10$ to maintain a high signal-to-noise ratio (S/N$>$3). We adopt two models to fit a surface density profile, i.e., (1) the pure S\'ersic model and (2) the S\'ersic plus exponential disk model.

The S\'ersic profile is given by \cite{1968adga.book.....S}
\begin{equation}
\Sigma(r) = \Sigma(0)\mathrm{e}^{-b_{n}(r/r_{e})^{1/n}},
\end{equation}
where $\Sigma(0)$ is the central surface density, $r_{e}$ is the effective radius, $n$ is the shape parameter and $b_{n}$ is a function of $n$, which is chosen to ensure that $r_{e}$ encloses half of the total mass.

The S\'ersic and exponential disk profile is given by \cite{1970ApJ...160..811F}
\begin{equation}
\Sigma(r) = \Sigma_{\mathrm{D}}(0)\mathrm{e}^{-r/h_{\mathrm{D}}} + \Sigma_{\mathrm{B}}(0)\mathrm{e}^{-b_{n}(r/r_{e\mathrm{B}})^{1/n_{\mathrm{B}}}},
\end{equation}
where $\Sigma_{\mathrm{D}}(0)$ is the central surface density of the disk, $h_{\mathrm{D}}$ is the disk scale length, and $\Sigma_{\mathrm{B}}(0)$, $r_{e\mathrm{B}}$, and $n_{\mathrm{B}}$ are the central surface density, the effective radius, and shape parameter of the bulge, respectively. Through these fittings, we obtain the effective radius of these two bulges (or a single bulge after the final merger of the two MBHs). Thus we can measure the bulge mass associated with each MBH (or the remnant MBH after merger) by integrating the bulge surface density profile.

The velocity dispersion ($\sigma_{*}$) of stars within the half-light radius ($r_{\rm e}$) of the galaxy bulge at a given time can be calculated (see \cite{2006ApJ...641...90R}, \cite{2006ApJ...641...21R}). In each snapshot, we select the stellar particles within $r_{\rm e}$ of a bulge and calculate the mass-weighted velocity dispersion as \cite{2012ApJ...747...33S}
\begin{equation}
\sigma_* = \sqrt{v_{i}^{2}m_{i}/M - (v_{i}m_{i}/M)^{2}},
\end{equation}
where 
\begin{equation}
M = \sum_{i} m_{i}.
\end{equation}
and $v_{i}$ is the line-of-sight (LOS) velocity of a selected stellar particle. The LOS velocity dispersion is calculated and the repeated indices in above equation represent the summation over all particles. We simply assume the stellar mass-to-light ratio ($M_*/L$) as constant for all stellar particles, which causes no difference between the light-weighted and mass-weighted velocity dispersions.
The calculated velocity dispersion may depend on viewing angle because some newly formed stars are present on a disk and some stars from the original disk may still maintain some of its initial angular momentum. We calculate the LOS velocity dispersion for $1,000$ randomly selected viewing angles, and take the typical LOS velocity dispersion as the average of those obtained from above $1,000$ viewing angles.

\subsection{MBH mass growth}

Both MBHs grow with time due to accretion during the merger (see Eq.~\ref{eq:Mgrow}).
At the late stage of the merger, the separation of the two MBHs oscillates around the softening length due to the spatial resolution limitation. We assume that these two MBHs coalesce immediately once their separation is smaller than the softening length and the remnant MBH mass is the the total mass of these two MBH right before their merger. Merger driven that the two MBHs may rapidly form a binary MBH (BBH) that is indistinguishable from a single MBH and/or the BBH quickly merge on a timescale that is substantially smaller than the galaxy merge timescale \cite{2007Sci...316.1874M, 2002MNRAS.331..935Y, 2018ApJ...868...97K}. We directly output the evolution of the MBH masses in each simulation.

\begin{tablehere}
%
\caption{Mass and spatial resolutions of different particles} 
\begin{center}

\begin{tabular}{ccc}
\hline
Particle & Mass ($M_{\odot}$) & Softening Length (pc)\\
\hline
Dark Matter & $1.1 \times 10^5$ & 30 \\
Bulge & $3.7 \times 10^{3}$ & 10 \\
Disk & $3.7 \times 10^{3}$ & 10 \\

Gas & $4.6 \times 10^{3}$ & 20 \\
\hline
\end{tabular}
\end{center}
\label{tab:ini_setup}

\end{tablehere}

\section{Results}
\label{sec:result}

We analyze data output from our simulation for each snapshot and obtain our main results as detailed below. 

\begin{figure*}[htp]
\includegraphics[width=0.99\textwidth]{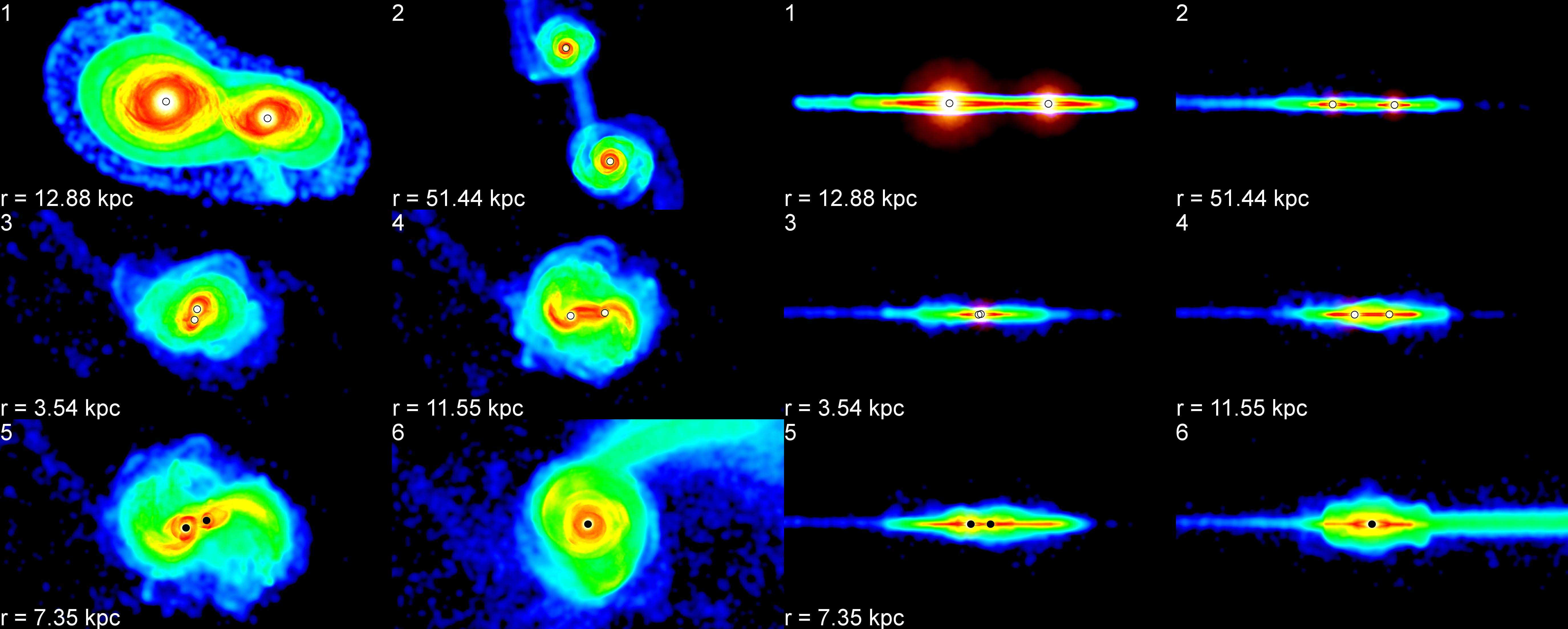}\\
\includegraphics[width=0.99\textwidth]{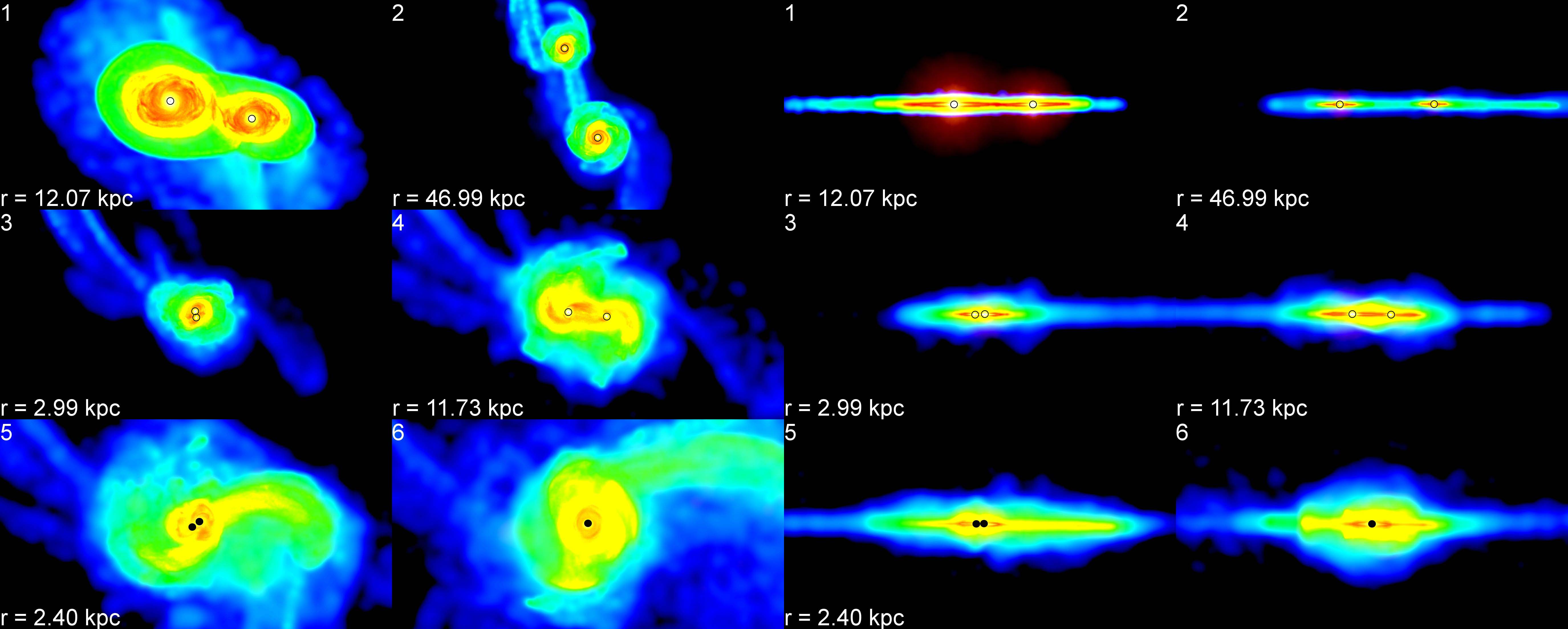}\\
\includegraphics[width=0.99\textwidth]{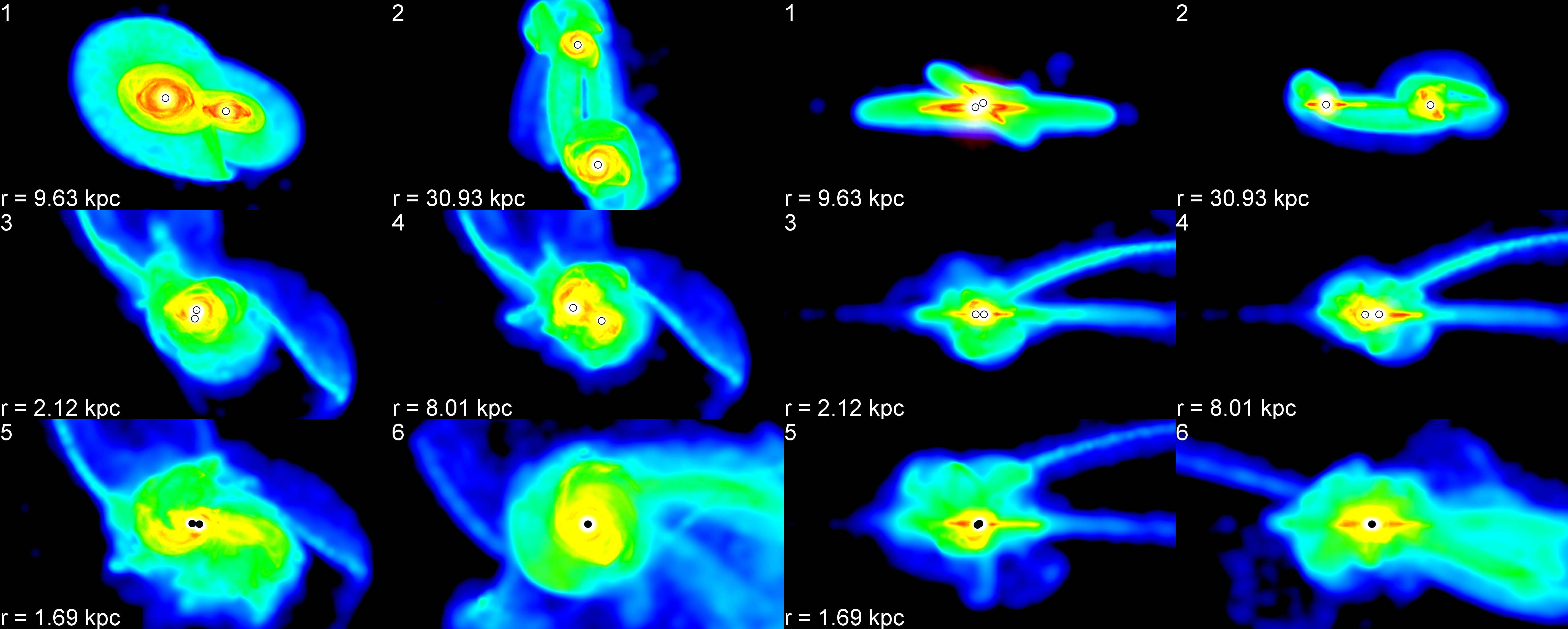}
\caption{Stellar density snapshots of the $z=2$, $i=0^{\circ}$ (top three rows), $z=3$, $i=0^{\circ}$ (middle three rows), and $z=3$, $i=45^{\circ}$ (bottom three rows) by viewing the primary galaxy in face-on (two left columns) and edge-on (two right column), respectively. The six snapshots (1-6 as labeled in the top left corner of each panel) of each merger correspond to the six stages as labeled in Table~\ref{tab:best_fit}: (1) the first pericentric passage, (2) the first apocentric passage after the first pericentric passage, (3) the second pericentric passage, (4) the second apocentric passage, (5) a time at the dAGN stage, and (6) the moment when the separation of two MBHs reaches the softening length. The length scales for all panels are the same, and the separation of the two MBHs is labeled at the bottom left in each panel. The positions of MBHs are marked with black open circles in snapshots (1)-(4), and black solid circles in snapshots (5) and (6). The size of each circle is arbitrary and not scaled to the real size of the nuclei/MBHs.} 
\label{fig:galaxy}
\end{figure*}

We first visualize six snapshots for each of the three simulations we performed in Figure~\ref{fig:galaxy}. The top, middle, and bottom rows of Figure~\ref{fig:galaxy} show the simulations with $z=2$ and $i=0^{\circ}$, $z=3$ and $i=0^{\circ}$, and $z=3$ and $i=45^{\circ}$, respectively. The two left and two right columns show those snapshots viewed at a direction perpendicular (face-on) or parallel to (edge-on) the disk plane of the primary galaxy, respectively. The six snapshots show the merging galaxy at the first pericentric passage (1), the first apocentric passage after the first pericentric passage (2), the second pericentric passage (3), the second apocentric passage (4), a dAGN stage (5), and the moment when the two MBHs coalesced (6). A tidal bridge is clearly seen after the first pericentric passage (snapshot 2), which is the channel for transporting materials between the two galaxies. Tidal tails can be easily identified in snapshots 3, 4, 5, and 6, which can be taken as evidence of galaxy collisions. In snapshots 1, 2, 3, and 4, the two black open circles mark the location of the two MBHs. In snapshot 5, dAGNs are triggered and the location of the two MBHs is marked by solid black circles. The MBHs merge with each other in snapshot 5, and the nuclear activity declines then but it may last on a significant level for a long time (see Fig.~\ref{fig:bhacc}).

\subsection{dAGN}
\begin{figure*}[htp]
\centering
\includegraphics[scale=0.50]{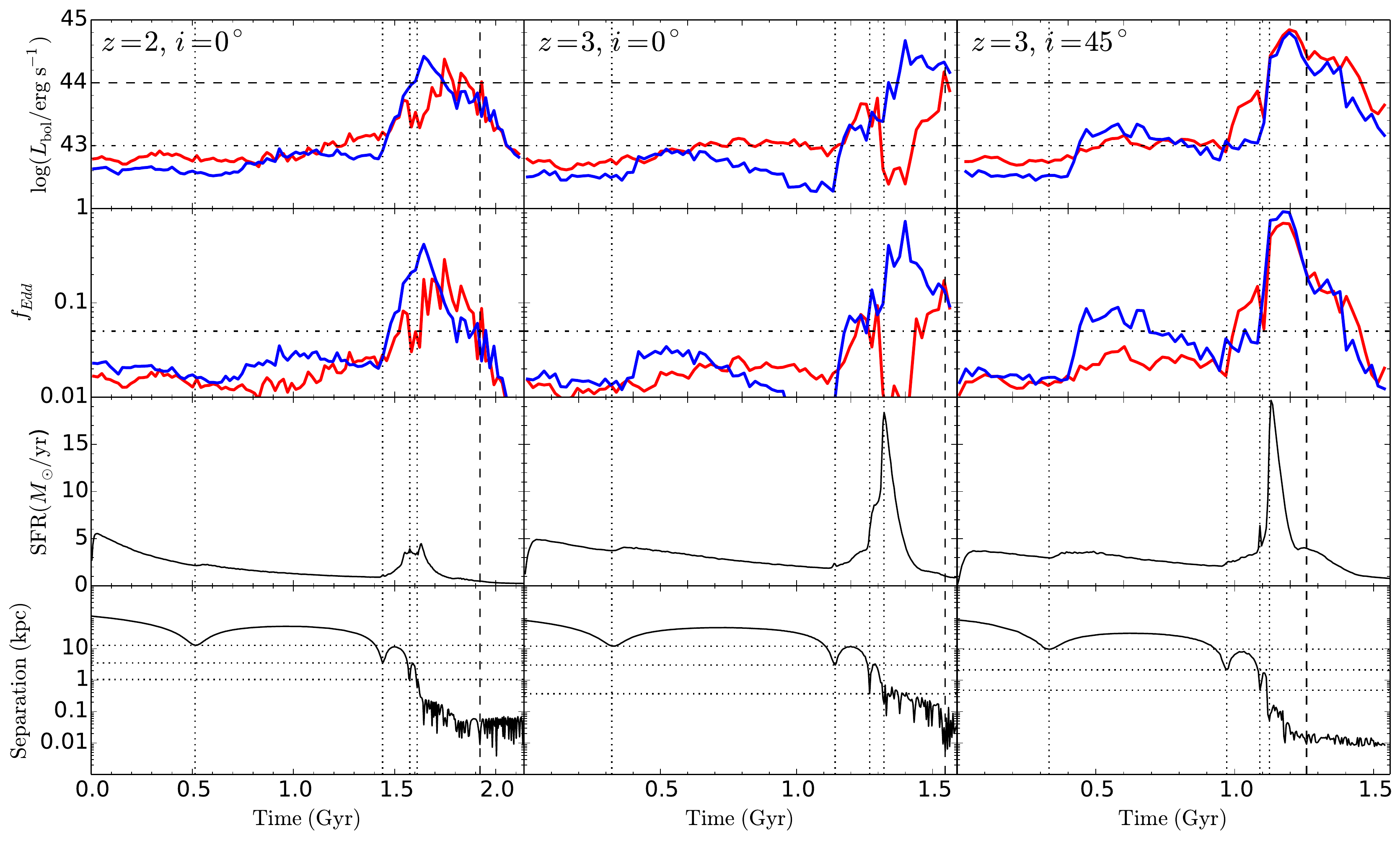}
\caption{Evolution of the nuclear luminosity, Eddington ratio, SFR, and the separation of two MBHs during the merging process of galaxy mergers. The left, middle, and right columns show the three simulated mergers starting from $z=2$ with an inclination angle of $i=0^{\circ}$ (left column), $z=3$ with $i=0^{\circ}$ (middle column), and $z=3$ with $i=45^{\circ}$ (right column), respectively. In the top two rows, the red and blue curves represent the bolometric luminosities (first row) and Eddington ratios (second row) of the two MBHs in the primary and secondary galaxies, respectively. Considering the difficulties of classifying the SFR of the two progenitor galaxies separately after their first pericentric passage, only the total SFR is shown in the third row panels. The bottom panel in each column shows the evolution of the separation of the MBHs. In each panel, the vertical dotted lines from left to right indicate the first, second, third, and fourth pericentric passages, respectively. The corresponding separation of the first three pericentric passages is marked by the horizontal dotted lines. The vertical dashed line in each panel of the bottom row marks the time when the separation of two MBHs first reaches the softening length. The horizontal dashed and dashed-dotted lines in each panel of the top row represent the luminosity thresholds $L_{\mathrm{bol}} = 10^{44}$ and $10^{43}$erg/s, respectively. The horizontal dashed-dotted lines in each panel of the second row represents the Eddington ratio threshold $f_{\mathrm{Edd}} = 0.05$.}
\label{fig:bhacc}
\end{figure*}

Figure~\ref{fig:bhacc} shows the evolution of the bolometric luminosity, Eddington ratio, host galaxy star formation, and the separation of the two MBHs. The left, middle, and right columns of Figure~\ref{fig:bhacc} show the simulation results for the galaxy mergers starting from $z = 2$ with an inclination angle of $i=0^{\circ}$, from $z = 3$ with $i=0^{\circ}$, and from $z=3$ with $i=45^{\circ}$, respectively.  

Our three simulations are different from each other in the sense of the nuclear activity triggering and the formation of dAGNs, according to the panels shown in Figure~\ref{fig:bhacc}. 
At the beginning of the merger, the SFR in the two progenitor galaxies decreases with elapsed time due to the gas consumption, while their nuclear activities are weak and the accretion rates are roughly constant, which means that the tidal torques have little effect on gas particles in the vicinity of each MBH and thus on feeding MBHs. After the first pericentric passage, the SFR is slightly enhanced in all three simulations. The tidal force begins to affect the gas accretion of the secondary MBH, which causes the Eddington ratio of the secondary MBH to exceed the primary one, and leads to a comparable bolometric luminosity in the two simulations with $i=0^{\circ}$. 

Once the merger goes through the second pericentric passage, the SFR is further enhanced and reaches its peak value at the third pericentric passage in the simulation with  $z=2$ (left column), while it climbs to a higher peak after the third pericentric passage in the other two simulations with $z=3$. Except for the different SFR behaviors in different mergers, the evolutions of the MBH activities in these three simulations are also different from each other.  
Considerable differences in the evolution of nuclear activities were the trigger in these three simulations. For the simulation with $z=2$, the increases in $f_{\rm Edd}$ and $L_{\rm bol}$ of the primary MBH are similar to those of the secondary MBH, but with a time lag of $\sim 0.1$\,Gyr. For the simulation with $z=3$ and $i=0^{\circ}$, the activity of the primary is significantly suppressed when it reaches the fourth pericentric passage MBH. $f_{\rm Edd}$ and $L_{\rm bol}$ decrease rapidly, which might due to the destruction and/or consumption of nuclear gas in the vicinity of the primary MBH as a result of the penetration of the nucleus of the secondary galaxy and/or rapid star formation. With the decreasing SFR, the activity of the primary MBH is enhanced and becomes comparable with that of the secondary MBH when their separation decreases to the scale of the softening length. For the simulation with $z=3$ and $i=45^{\circ}$, the activity of the primary MBH is similar to that of the secondary one after the third pericentric passage, for which the nuclear gas in the vicinity of the primary MBH is not affected substantially by the inclined collision of the two nuclei.

According to Figure~\ref{fig:bhacc}, a period exists in which both MBHs are active with comparably high bolometric luminosities, i.e., dAGN appears (see also snapshot 5 in Fig.~\ref{fig:galaxy}). As seen in Figure~\ref{fig:bhacc}, gas-rich merging galaxies appear as dAGNs only after the second or third pericentric passage (with a separation of a few kpc, comparable with the half light radius of the host galaxies), where both nuclei are significantly perturbed by the companion galaxy. This result confirms the conjecture made in the phenomenological model for dAGNs by  \cite{2011ApJ...738...92Y}.

The dAGN durations and fractions may be estimated by adopting different $L_{\rm bol}$, $f_{\rm Edd}$, and separation thresholds, which are listed in Table~\ref{tab:fagn}. If a lower luminosity threshold ($L_{\rm bol}>10^{43}$\,erg/s) and a small separation threshold ($>0.1$\,kpc) are adopted, dAGNs occur for a period $\gtrsim 0.2$\,Gyr and may appear in a significant time fraction ($\sim 18\%$ for the simulation with $i=0^{\circ}$ to $\sim 40\%$ for the simulation with $i=45^{\circ}$) of the whole merger period. However, if the detection capability and exposure time requirement of current telescopes are taken into account, then only those dAGNs with a larger separation or higher luminosity can be detected \cite{2009ApJ...693.1554F}. If $L_{\rm bol}>10^{44}$erg/s is set as the luminosity threshold, dAGNs can be observed in only a fraction of $\sim 5\%$ of the whole merger period for a separation threshold of $>0.1$\,kpc, and are almost undetectable for a separation threshold of $>1$kpc. If the $f_{\rm Edd}>0.05$ is set as detection threshold, then we can observe dAGNs with separation $>0.1$\,kpc at a  fraction of $10\%$-$15\%$ of the whole merger period for those simulations with $i=0^{\circ}$, and this fraction decreases with increasing inclination angle because both MBHs have significant accretion that occurs only at small separations (see right panels of Fig.~\ref{fig:bhacc}). If we can only resolve those dual cores with separation $>1$\,kpc, then the detection rate decreases to $3\%$-$7\%$ for those $i=0^{\circ}$ mergers, and only $2\%$ for the $i=45^{\circ}$ merger. In those 1:2 spiral-spiral galaxy mergers, the detection rate are all close to zero by adopting the threshold of $f_{\rm Edd}>0.1$ and separation $>1$\,kpc  mainly because high Eddington ratio accretion occurs at a small separation only.

For dAGNs in different simulations with an adopted threshold of $L_{\rm bol}>10^{43}$erg/s or $f_{\rm Edd}>0.05$, the duration time is roughly consistent, ranging from $\sim 45-160$\,Myr \cite{2017MNRAS.469.4437C} to $\sim 180-250$ \,Myr \cite{2012ApJ...748L...7V} in the spiral-spiral major merger cases, which are consistent with ours (for the $z = 3, i = 0^{\circ}$ case, $\sim 110-280$\,Myr). We also find that dAGNs commonly emerge at small separations ($\sim$\,kpc), thereby suggesting that galaxy mergers can trigger kpc-scale dAGNs \cite{2012ApJ...746L..22K, 2018ApJ...867...66C, 2013ApJ...777...64C, 2015ApJ...806..219C}.

\begin{tablehere}
\caption{Dual AGN fractions versus different $L_{\rm bol}$, $f_{\rm Edd}$, and separation thresholds.}
\centering

\resizebox{87mm}{25mm}
{
\begin{tabular}{cccccc}
\hline
\multirow{2}{*}{Simulation} & \multirow{2}{*}{Threshold} & \multicolumn{2}{c}{$r > 1$ kpc} & \multicolumn{2}{c}{$r > 0.1$ kpc}\\
& & $t_{\rm dAGN}$ & $t_{\rm dAGN}/t_{\rm sim}$ & $t_{\rm dAGN}$ & $t_{\rm dAGN}/t_{\rm sim}$\\
\hline

$z = 2$ & $L_{\rm bol} > 10^{43}$ & 0.26  & 0.12 & 0.39 & 0.18\\
$i=0^{\circ}$ & $L_{\rm bol} > 10^{44}$ & 0 &  0 & 0.10 & 0.05\\
& $f_{\rm Edd} > 0.05$ & 0.07 & 0.03 & 0.21 & 0.10\\
& $f_{\rm Edd} > 0.1$ & 0.02 & 0.01 & 0.13 & 0.06\\
%

$z = 3$ & $L_{\rm bol} > 10^{43}$ & 0.15 & 0.10 & 0.28 & 0.18\\
$i=0^{\circ}$ & $L_{\rm bol} > 10^{44}$ & 0.01 & 0.01 & 0.07 & 0.05\\
& $f_{\rm Edd} > 0.05$ & 0.11 & 0.07 & 0.24 & 0.15\\
& $f_{\rm Edd} > 0.1$ & 0.01 & 0.01 & 0.08 & 0.05\\
%

$z = 3$ & $L_{\rm bol} > 10^{43}$ & 0.14 & 0.09 & 0.21 & 0.4\\
$i=45^{\circ}$ & $L_{\rm bol} > 10^{44}$ & 0 & 0 & 0.06 & 0.04\\
& $f_{\rm Edd} > 0.05$ & 0.03 & 0.02 & 0.10 & 0.07\\
& $f_{\rm Edd} > 0.1$ & 0 & 0 & 0.06 & 0.04\\
\hline
\end{tabular}
}
\label{tab:fagn}
\end{tablehere}

 We calculate the averaged column density of gas along the LOS with 10 different viewing angles from edge-on to face-on and find that the column density of gas never reachs the Compton thick regime ($m_{\rm H} = 10^{24}\ \mathrm{cm}^{-2}$) in the dAGN stages when the separation $r>0.1$ kpc. At the late stage of the merger (when the separation $r < 0.1$\,kpc), the gas column density can increase to $\sim 1.3 \times 10^{24}\ \mathrm{cm}^{-2}$. This result is consistent with the simulation by \cite{2018MNRAS.478.3056B}, in which claimed that the gas column density can reach to the Compton thick regime in the late stage of the galaxy merger and dAGNs are obscured, but these dAGN systems can be identified by the mid-infrared color selection.

\begin{table*}
\begin{center}

\caption{Recovered parameters derived from fittings of the pure S\'ersic and S\'ersic + Exponential disk profiles} 

\begin{tabular}{clcccccccccc}
\hline
\multicolumn{2}{c}{} & \multicolumn{4}{c}{S\'ersic Profile} & \multicolumn{6}{c}{S\'ersic + exponential Disk}\\
 \cmidrule(r){3-6}  \cmidrule(r){7-12}
\multirow{2}{*}{} & Time (Gyr) & \multicolumn{2}{c}{Primary Galaxy} & \multicolumn{2}{c}{Secondary Galaxy} & \multicolumn{3}{c}{Primary Galaxy} & \multicolumn{3}{c}{Secondary Galaxy}\\
\cmidrule(r){3-4} \cmidrule(r){5-6} \cmidrule(r){7-9} \cmidrule(r){10-12}
& & $n$ & $r_{e}(\mathrm{kpc})$ & $n$ & $r_{e}(\mathrm{kpc})$ & $n_{\mathrm{B}}$ & $r_{e\mathrm{B}}(\mathrm{kpc})$ & $h_{\mathrm{D}}(\mathrm{kpc})$ & $n_{\mathrm{B}}$ & $r_{e\mathrm{B}}(\mathrm{kpc})$ & $h_{\mathrm{D}}(\mathrm{kpc})$\\
\hline
%
$z=2,\ i=0^{\circ}$ & 0.00     & 2.84 & 1.04 & 2.82 & 0.98 & 2.98 & 0.74 & 1.20 & 2.76 & 0.52 & 0.92\\
& 0.40     & 2.32 & 1.29 & 2.05 & 1.27 & 2.40 & 0.98 & 1.02 & 1.64 & 0.51 & 0.70\\
& 0.52 (1) & 1.99 & 0.83 & 2.21 & 1.40 & 2.56 & 0.96 & 0.85 & 4.14 & 0.75 & 0.54\\
& 0.80     & 2.03 & 0.83 & 2.17 & 1.71 & 2.49 & 1.00 & 0.91 & 2.80 & 0.84 & 0.66\\
& 0.98 (2) & 2.08 & 2.20 & 2.33 & 1.42 & 1.60 & 0.97 & 0.81 & 3.65 & 0.91 & 0.58\\
& 1.20     & 2.06 & 1.44 & 2.15 & 1.49 & 1.78 & 1.11 & 0.88 & 2.73 & 0.89 & 0.65\\
& 1.44 (3) & 2.15 & 0.72 & 2.15 & 0.88 & 2.24 & 1.17 & 1.04 & 2.75 & 1.01 & 0.73\\
& 1.49 (4) & 2.44 & 0.73 & 2.62 & 3.59 & 2.11 & 1.23 & 3.00 & 1.58 & 0.99 & 3.06\\
& 1.55 (5) & 2.31 & 1.68 & 2.98 & 1.23 & 1.76 & 1.34 & 1.02 & 3.43 & 0.97 & 0.62\\
& 1.60     & 2.17 & 1.73 & 3.52 & 1.90 & 1.70 & 1.34 & 1.81 & 3.12 & 1.10 & 1.93\\
& 1.92 (6) & 3.09 & 1.18 & 3.09 & 1.18 & 2.84 & 1.25 & 1.73 & 2.84 & 1.25 & 1.73\\
& 2.00     & 3.04 & 1.07 & 3.04 & 1.07 & 2.64 & 1.54 & 1.85 & 2.64 & 1.54 & 1.85\\
& 2.13     & 3.13 & 1.27 & 3.13 & 1.27 & 2.87 & 1.61 & 2.05 & 2.87 & 1.61 & 2.05\\
\hline
%
$z=3,\ i=0^{\circ}$ & 0.00     & 4.25 & 0.70 & 3.13 & 1.25 & 2.75 & 0.75 & 1.53 & 3.11 & 0.54 & 1.25\\
& 0.33 (1) & 2.16 & 2.74 & 2.29 & 1.53 & 2.97 & 0.86 & 0.87 & 4.11 & 0.62 & 0.59\\
& 0.40     & 1.79 & 2.17 & 1.94 & 1.42 & 1.50 & 1.00 & 0.76 & 1.56 & 0.75 & 0.59\\
& 0.73 (2) & 2.19 & 2.38 & 2.27 & 1.55 & 1.87 & 1.03 & 1.03 & 2.68 & 0.77 & 0.77\\
& 0.80     & 2.13 & 2.65 & 2.07 & 0.91 & 2.54 & 1.05 & 0.99 & 1.79 & 0.80 & 0.86\\
& 1.14 (3) & 2.15 & 4.76 & 2.16 & 4.78 & 2.40 & 1.16 & 0.89 & 1.40 & 0.95 & 1.43\\
& 1.19 (4) & 2.05 & 1.31 & 2.25 & 0.76 & 2.51 & 1.16 & 0.67 & 2.02 & 0.97 & 1.51\\
& 1.20     & 2.22 & 1.26 & 1.88 & 1.06 & 1.95 & 1.22 & 1.15 & 1.71 & 1.08 & 2.63\\
& 1.30 (5) & 2.48 & 0.44 & 3.23 & 2.84 & 1.43 & 1.22 & 1.29 & 1.27 & 1.14 & 2.09\\
& 1.33     & 1.48 & 1.73 & 2.11 & 1.98 & 2.53 & 1.25 & 0.98 & 1.62 & 1.29 & 1.27\\
& 1.40 (*)  & 3.32 & 0.49 & 3.37 & 1.94 & 2.07 & 1.23 & 1.58 & 1.31 & 1.29 & 1.52\\
& 1.54 (6) & 3.62 & 1.32 & 3.62 & 1.32 & 3.04 & 1.29 & 1.64 & 3.04 & 1.29 & 1.64\\
& 1.60     & 3.36 & 1.30 & 3.36 & 1.30 & 3.36 & 1.39 & 2.23 & 3.36 & 1.39 & 2.23\\
& 1.74     & 3.37 & 1.95 & 3.37 & 1.95 & 2.98 & 1.62 & 1.69 & 2.98 & 1.62 & 1.69\\
\hline
%
$z=3,\ i=45^{\circ}$ & 0.00     & 4.25 & 0.70 & 3.57 & 0.67 & 2.75 & 0.75 & 1.53 & 2.98 & 0.62 & 1.01\\
& 0.33 (1) & 2.07 & 0.57 & 2.37 & 0.73 & 2.75 & 0.69 & 0.84 & 2.39 & 0.79 & 0.90\\
& 0.40     & 2.48 & 0.94 & 2.01 & 0.66 & 3.10 & 0.95 & 0.75 & 1.56 & 0.75 & 0.62\\
& 0.64 (2) & 2.44 & 1.14 & 2.60 & 0.76 & 2.59 & 1.00 & 1.14 & 2.01 & 0.87 & 1.14\\
& 0.80     & 2.43 & 1.19 & 2.57 & 0.71 & 2.57 & 1.06 & 1.03 & 1.87 & 0.87 & 1.03\\
& 0.97 (3) & 2.40 & 1.29 & 2.17 & 0.77 & 2.79 & 1.11 & 0.96 & 1.83 & 0.89 & 1.23\\
& 1.02 (4) & 2.11 & 0.87 & 3.18 & 0.76 & 1.92 & 1.12 & 1.42 & 1.75 & 1.14 & 1.22\\
& 1.10 (5) & 1.59 & 0.90 & 2.38 & 1.38 & 1.36 & 1.28 & 1.87 & 2.48 & 1.18 & 1.08\\
& 1.20     & 3.64 & 0.91 & 2.99 & 0.91 & 2.41 & 1.86 & 1.89 & 2.03 & 1.43 & 1.94\\
& 1.25 (6) & 3.46 & 0.98 & 3.46 & 0.97 & 3.60 & 2.57 & 0.92 & 3.60 & 2.57 & 0.92\\
& 1.56     & 3.67 & 1.08 & 3.67 & 1.08 & 3.96 & 2.02 & 0.65 & 3.96 & 2.02 & 0.65\\
\hline
\end{tabular}
\label{tab:best_fit}
\end{center}
Note: The numbers (1)-(6) in the brackets in the Time column show the time at the six stages as shown in Fig.~\ref{fig:galaxy}. For the $z=3$, $i=0^{\circ}$ simulation, we use the symbol (*) to represent the time when the black hole mass of the secondary galaxy is equal to that of the primary galaxy.

\end{table*}

\subsection{Evolution of $M_{\bullet}$-$M_{\mathrm{b}}$ relation}
\label{subsec:mmbulge}
\begin{figure}[H]
\centering
\includegraphics[width=0.49\textwidth]{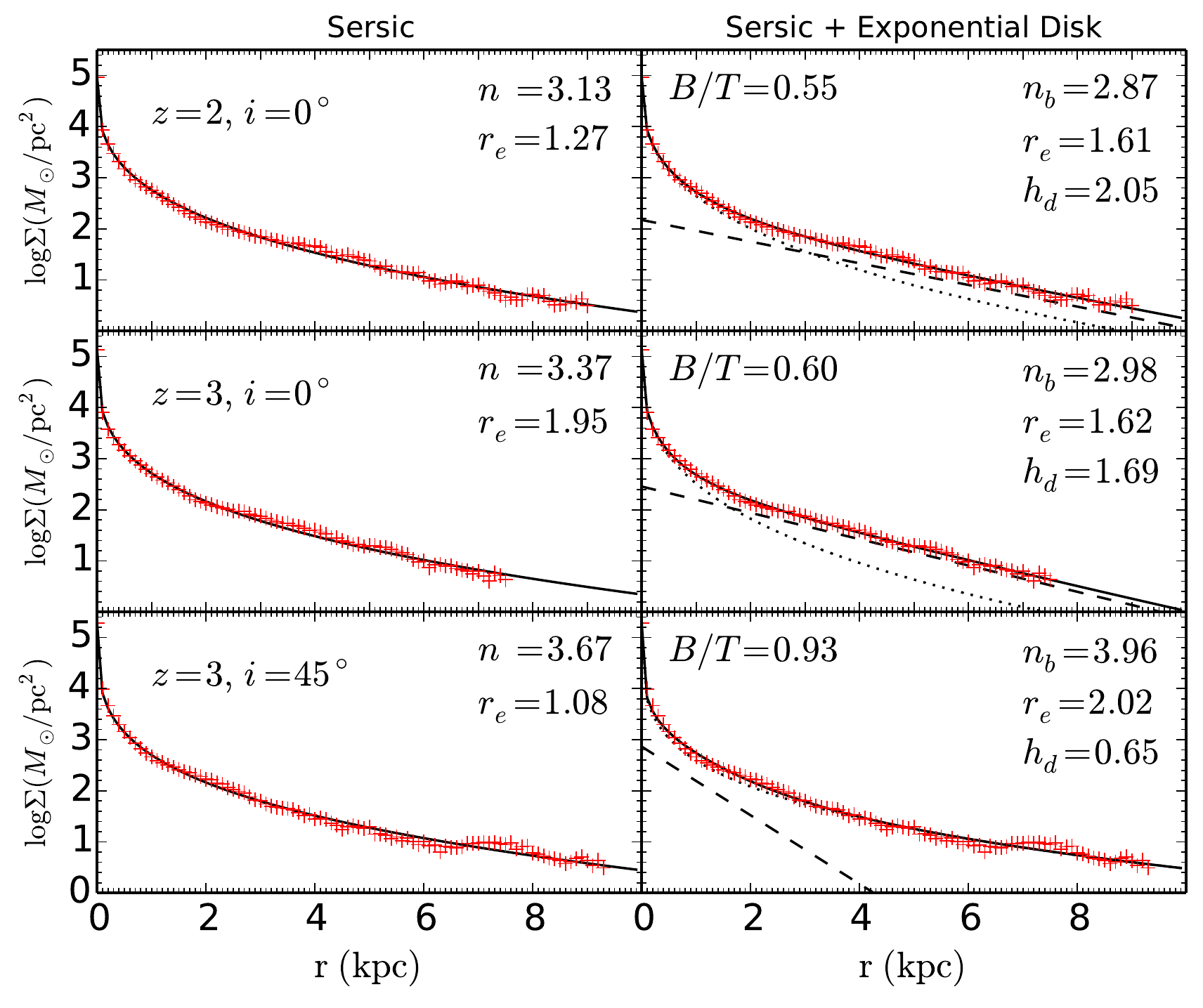}
\caption{Stellar surface density profile fittings of the final merged galaxies in the three simulations (from top to bottom rows). The pure Sersic profile fittings are shown in the left column, and the Sersic + exponential disk fittings are shown in the right column. In each panel, the red profile is the stellar density profile of the simulation, and the black solid line shows the model fitting. For the Sersic + exponential disk fitting, the dotted line shows the Sersic component, and the dashed line presents the disk component. Those fitted parameters labeled in each panel are also listed in Table \ref{tab:best_fit}.}

\label{fig:fit_final}
\end{figure}

We analyze the morphology of merged galaxy to interpret the final product of the galaxy merger. Figure~\ref{fig:fit_final} shows the stellar density profile fittings of the three simulations, in which the Sersic + exponential disk model better fits the merged galaxies resulting from the simulations with ($z=2, i=0^{\circ}$) and ($z=3, i=0^{\circ}$) than the pure Sersic model, while the pure Sersic model better fits the merged galaxy of the simulation with ($z=3$, $i=45^{\circ}$) than the Sersic + exponential disk model. These results indicate that the final merged galaxies in the case ($z=2, i=0^{\circ}$) and ($z=3, i=0^{\circ}$) have their B/T$\sim0.6$, which are more like disk galaxies, while the product of the $z=3, i=45^{\circ}$ simulation is an elliptical galaxy. Those fitted parameters labeled in each panel are also listed in Table \ref{tab:best_fit}.

\begin{figure}[H]
\includegraphics[width=0.46\textwidth]{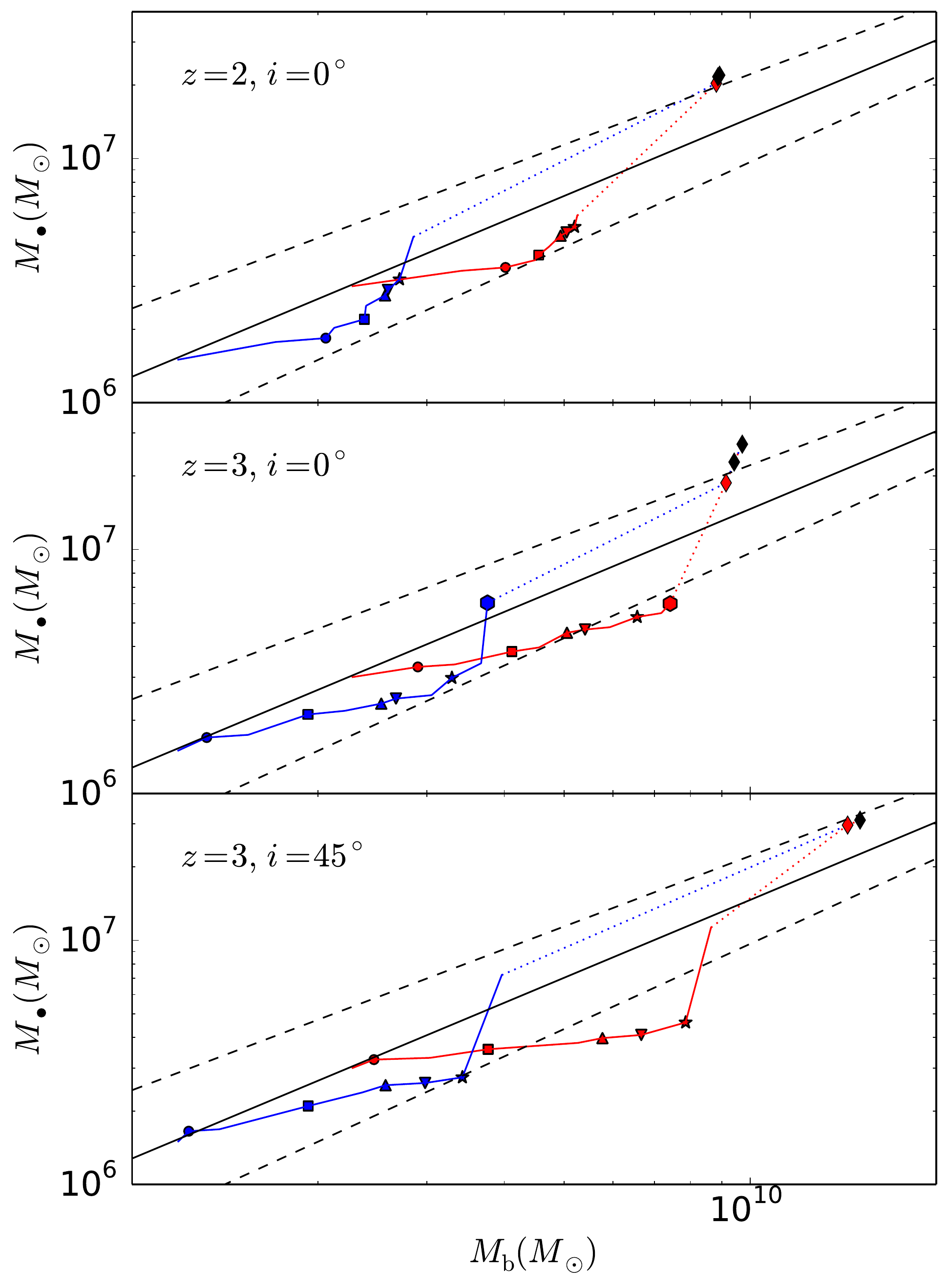}
\caption{
Evolution of $M_{\bullet}-M_{\mathrm{b}}$ relation. The three panels from top to bottom show the $z=2$ with $i=0^{\circ}$, $z=3$ with $i=0^{\circ}$, and $z=3$ with $i=45^{\circ}$ simulations, respectively. In each panel, the red line shows the evolution trend of the primary galaxy with the time stages listed in Table \ref{tab:best_fit}, the blue line shows that of the secondary galaxy. The dotted line is used to combine the point before the coalescence of BHs and the point at which the two BHs coalesced. The solid symbols on each line correspond to stages (1)-(6) with their images shown in Fig.~\ref{fig:galaxy}: circle for stage (1), box for stage (2), triangle for stage (3), inverted triangle for stage (4), star for stage (5), and diamond for stage (6). In the middle panel, the hexagon symbols represent the time when the two MBHs have equal mass, after which the MBH of the secondary galaxy is larger than that of the primary galaxy. In each panel, the black solid line represent the observed relation by \cite{2003ApJ...589L..21M}, and the two black dashed lines represents the $1\sigma$ uncertainty of the relation.}
\label{fig:mmbulge}
\end{figure}

The evolution of bulge and its mass ($M_{\rm b}$) and velocity dispersion ($\sigma_*$) can be obtained according to the procedures described in Section~\ref{sec:methods} and the mass growth of MBHs can also be output from the simulations. Figure~\ref{fig:mmbulge} shows the evolution of MBHs on the $M_{\bullet}$-$M_{\mathrm{b}}$ plane for the simulation with ($z=2, i=0^{\circ}$), ($z=3, i=0^{\circ}$), and ($z=3, i=45^{\circ}$). On the basis of our initial parameter setup, the masses of MBHs and bulges follow the $M_{\bullet}$-$M_{\mathrm{b}}$ relation \cite{2003ApJ...589L..21M}. During the merging process, the secondary MBH is first triggered and grows faster than the primary MBH in all three simulations because the tidal torque from the primary galaxy (with a larger mass and heavier nucleus) first perturbs the secondary nucleus and lead to the sink of gas into the vicinity of the secondary MBH. In the simulation with $z=3$ and $i=0^{\circ}$, the mass of the secondary MBH can even grow up to a mass larger than that of the primary MBH at the time labeled with the hexagon symbol.

Before the two MBHs merged together, galaxy bulges always grow faster than MBHs or predate the growth of MBHs if regulated with the $M_{\bullet}$-$M_{\mathrm{b}}$ relation, which is due to the high SFR but relatively low $f_{\rm Edd}$. Once the two MBHs merged together, the mass increase of the merged MBH slightly exceeds the value predicted by the $M_{\bullet}$-$M_{\mathrm{b}}$ relation in all the three simulations.

\subsection{Evolution of the $M_{\bullet}-\sigma_{*}$ relation}
\label{subsec:msigma}
Given that fractions of different stellar particles inside the effective radius of the bulge have large variation from the beginning to the end of the merger, the stellar velocity dispersion $\sigma_*$ should also vary following the merger process. With the current simulations, we can determine whether the newly joined stellar particles in the disk, or the orbital angular momentum of the merging galaxies is the dominant reason of $M_{\bullet}-\sigma_*$ evolution.

\begin{figure*}[htp]
\includegraphics[scale=0.54]{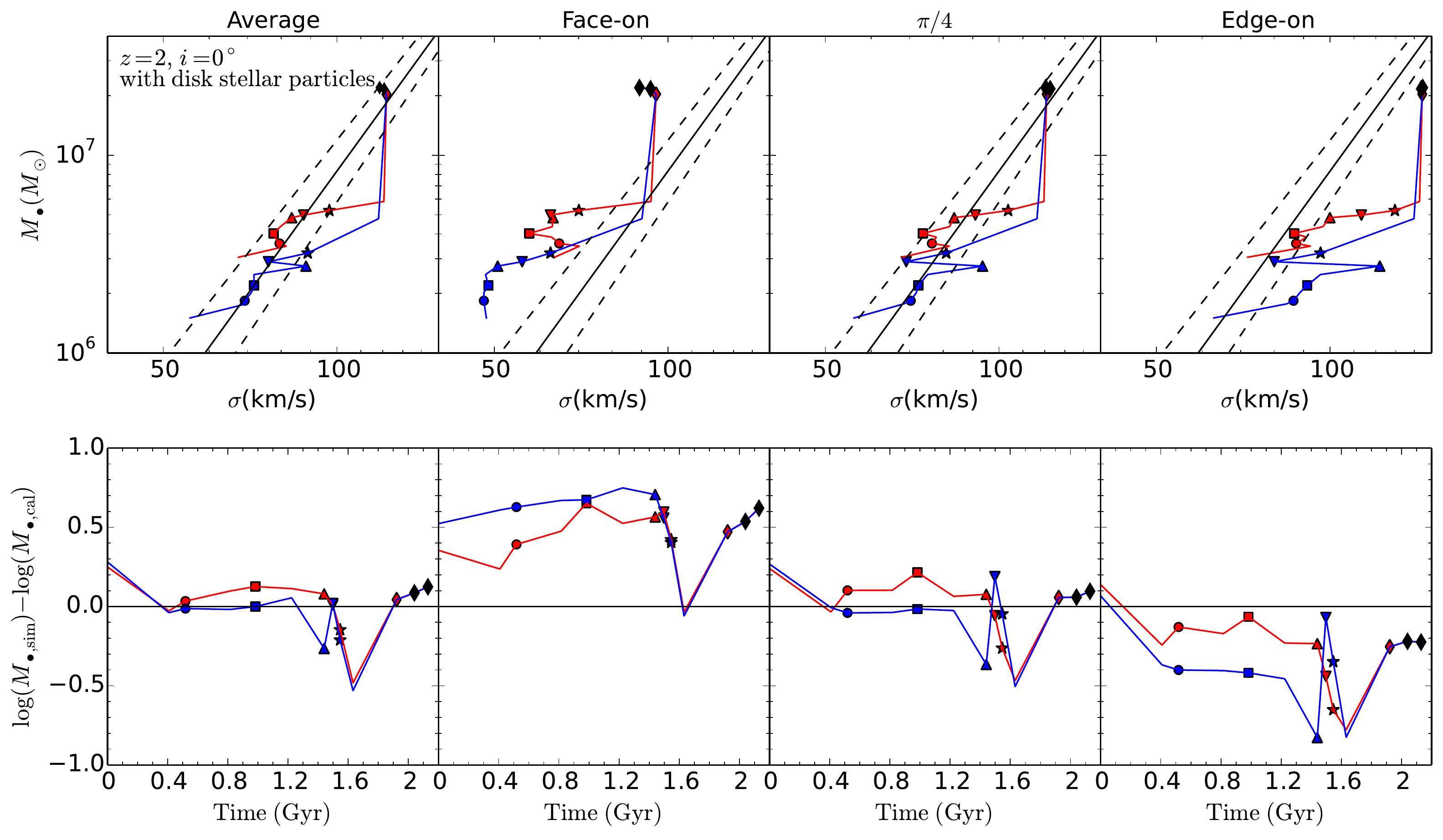}
\caption{Evolution of $M_{\bullet}$-$\sigma_{*}$ relation in $z = 2, i=0^{\circ}$ simulation. Here $\sigma_*$ include all the stellar particles inside the effective radius of the bulge. The top four panels from left to right show the averaged $\sigma_*$, $\sigma_*$ viewed from face-on, $45^{\circ}$, and edge-on, respectively. The bottom four panels show the corresponding differences between the black hole mass in the simulation ($M_{\mathrm{BH,\ sim}}$) and the black hole mass calculated based on the averaged $\sigma_*$ ($M_{\mathrm{BH,\ cal}}$). In each panel, the colored lines and symbols are the same as shown in Fig. \ref{fig:mmbulge}. The black solid lines are the observed $M_{\bullet}$-$\sigma_{*}$ relation by \cite{2002ApJ...574..740T} and the black dashed lines show the $1\sigma$ uncertainty in the relation. Here the averaged $\sigma_*$ is calculated based on 1000 random orientations.}
\label{fig:msigma_z2_d}
\end{figure*}

Figure~\ref{fig:msigma_z2_d} shows the evolution of the $M_{\bullet}-\sigma_*$ during the merging of galaxies for the simulation with $z=2$ and $i=0^{\circ}$. From the left to right, we show the results with $\sigma_*$ obtained by averaging over different viewing angles (average), viewing at a direction perpendicular to (face-on), inclined by an angle of $45^{\circ}$ to (inclined), and parallel to (edge-on) the disk plane of the primary galaxy, respectively. In the top four panels, we also show the observationally determined $M_{\bullet}-\sigma_*$ relation for comparison, i.e., $\log(M_{\bullet}/M_{\odot}) = a + b \log(\sigma/\sigma_{0})$, where $a = 8.13 \pm 0.06,\ b = 4.02 \pm 0.32,\ \sigma_{0} = 200\mathrm{km/s}$ \cite{2002ApJ...574..740T}. The differences between the MBH mass in the simulation ($M_{\mathrm{BH,\ sim}}$) and that predicted by the observed $M_{\bullet}$-$\sigma_{*}$ relation ($M_{\mathrm{BH,\ cal}}$) are correspondingly shown in the bottom panels. 

As described in Section 2.1, the relation between the MBH and bulge masses follows the $M_{\bullet}$-$M_{\rm b}$ relation, and the spatial mass density distribution follows the Hernquist profile. These initial setup causes a lower $\sigma_*$ compared with the above $M_{\bullet}$-$\sigma_{*}$ relation at the beginning as shown in the top-left panel of Figure~\ref{fig:msigma_z2_d}. Thsi results may be due to the fact that the system may not well relax at that time. The solid red (primary galaxy) and blue (secondary galaxy) circles correspond to the first pericentric passage with a separation of $12.88$\,kpc for this $z=2, i=0^{\circ}$ simulation, which shows that $\sigma_*$ increases with time and finally matches the velocity dispersion predicted by the observed relation before or around the first pericentric passage. Before the dAGN stage, the averaged $\sigma_*$ oscillates around the observed $M_{\bullet}$-$\sigma_{*}$ relation and has no large deviation. After the dAGN stage, the two galaxies are close and the two bulges are destroyed violently, thereby causing the averaged $\sigma_*$ to be larger than that predicted by the relation. When the two MBHs merged together, the new galaxy is built and the averaged $\sigma_*$ decreases after the relaxation in several hundreds of Myr. The $M_{\bullet}$-$\sigma_*$ relation in the simulation finally reverts to the observed relation.

The evolution of $M_{\bullet}$-$\sigma_*$ relation also varies with different viewing angles. The right three columns of Figure~\ref{fig:msigma_z2_d} show that the simulated $\sigma_*$ increases when the merging galaxies are viewed from face-on to edge-on perspective. In the face-on case, those stellar particles in the disk contribute little to the velocity dispersion, which then results in a smaller $\sigma_*$ than the averaged one. In the edge-on case, the rotation velocity of the disk particles are substantially larger than the velocity of bulge particles and thus leads to the $\sigma_*$ in this view angle being larger than the averaged one. The differences can be seen more clearly in the corresponding bottom panels of Figure~\ref{fig:msigma_z2_d}.

\begin{figure*}[htp]
\includegraphics[scale=0.58]{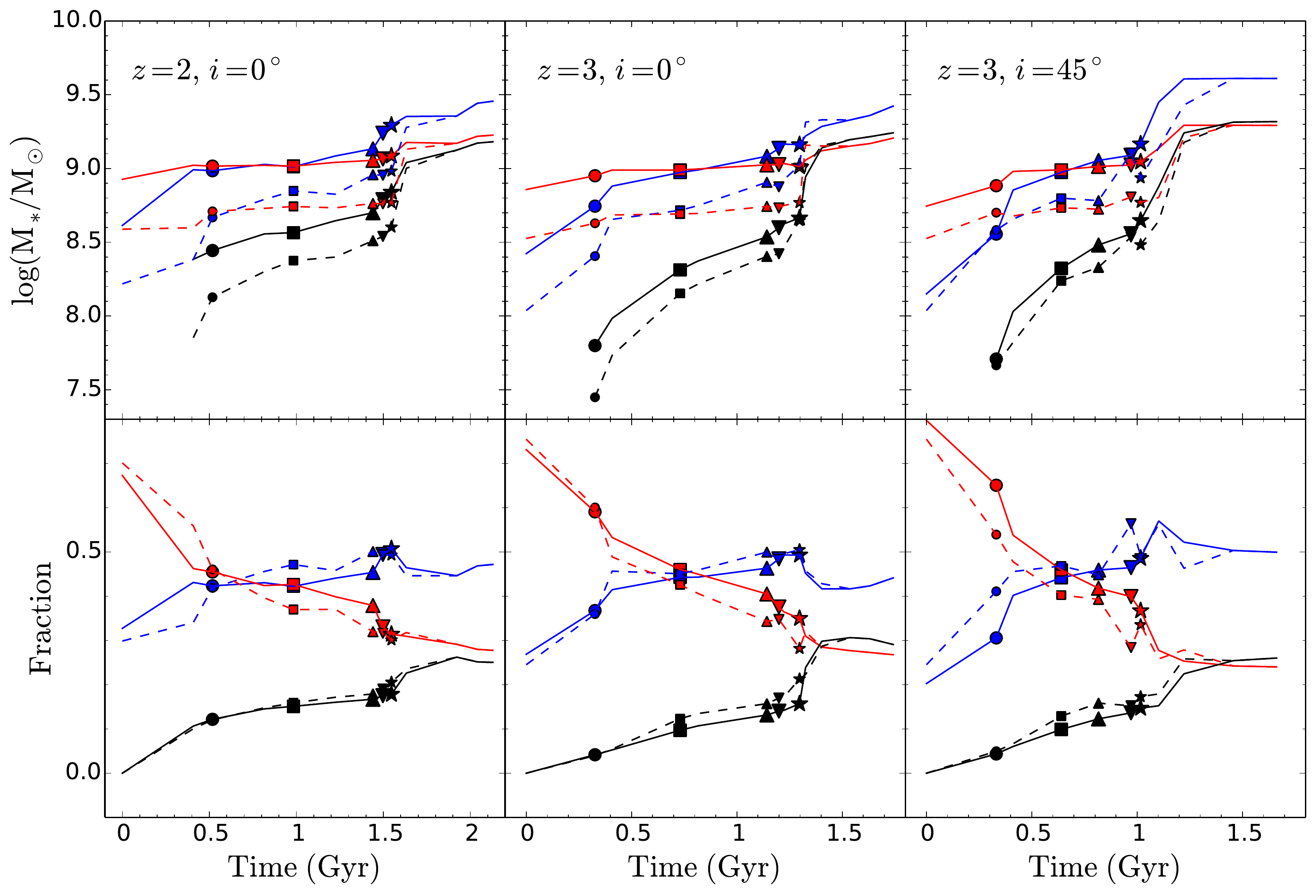}
\caption{Evolution of the three kinds of stellar particles inside the effective radius of the bulge. The stars that belong to the initial bulge of the progenitor galaxy are shown in red lines and symbols, stars that originate from the initial disk of the progenitor galaxy are shown in blue lines and symbols, and stars that are newly formed by those gas in the disk region are shown in black lines and symbols. The top three panels show the stellar mass evolution of the three components in the three simulations, and the bottom three panels show the corresponding mass fractions of the three stellar components. The symbols in different shapes are the same as those shown in Fig.~\ref{fig:mmbulge}.}
\label{fig:mfrac}
\end{figure*}

\begin{figure*}
\includegraphics[scale=0.54]{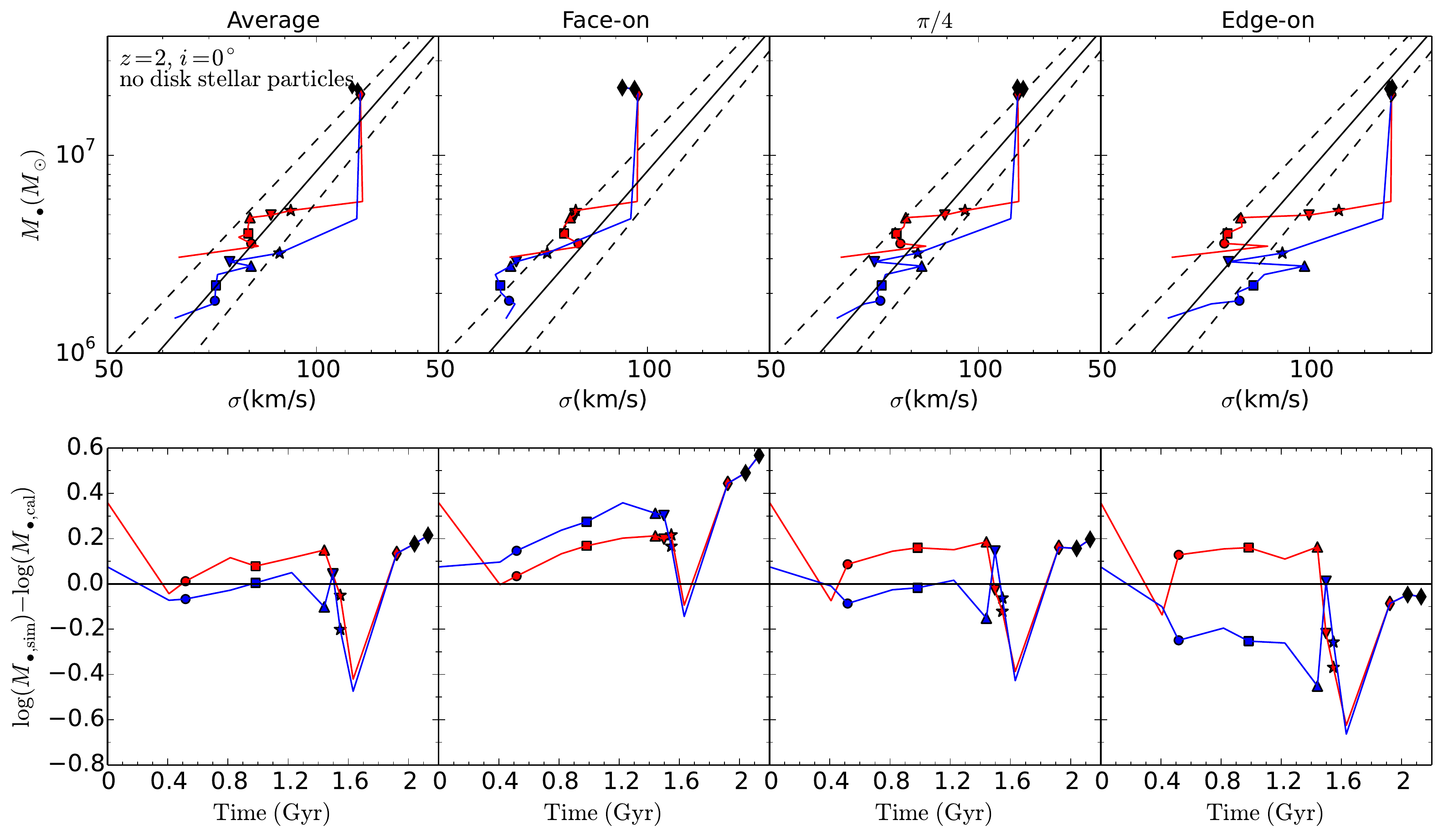}
\caption{The same plots as shown in Fig.~\ref{fig:msigma_z2_d}, but with $\sigma_*$ calculated only by excluding those stellar particles in initial disks of the progenitor galaxies.}
\label{fig:msigma_z2}
\end{figure*}

Figure~\ref{fig:mfrac} indicates that the stars in the progenitor disk and the newly formed ones take a total fraction of $75\%$ of the final bulge, which means that $\sigma_*$ might have a large variation if the galaxy is viewed from different inclination angles. To investigate whether the stellar particles in the initial disk dominate the $\sigma_*$ evolution, we exclude these particles and re-plot the results in Figure~\ref{fig:msigma_z2}. By comparing Figures~\ref{fig:msigma_z2_d} and \ref{fig:msigma_z2}, we can see that the basic evolution trends are similar, although the exact values of $\sigma_*$ have some differences. This finding proves that the evolution trend should not be due to the newly added disk particles, but to the orbital angular momentum of the two merging galaxies. As we can also see from Figure~\ref{fig:galaxy}, those merged galaxies are all quite flat, even in the case $z=0, i=45^\circ$, which suggests the angular momentum of the two merging galaxies are important in supporting the system. 

\begin{figure*}[htp]
\includegraphics[width=0.99\textwidth]{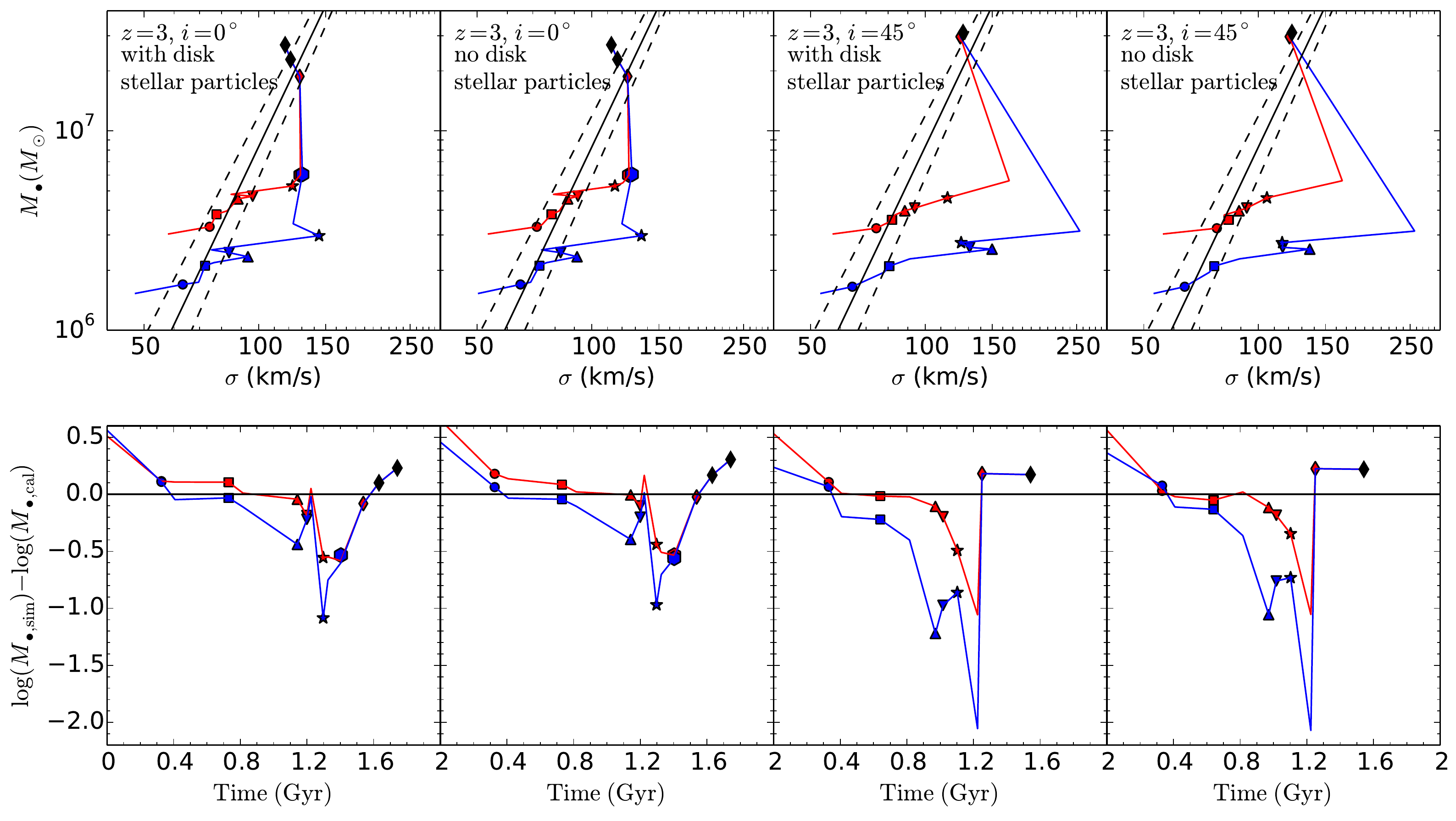}
\caption{Evolution of the $M_{\bullet}$-$\sigma_{*}$ relation in $z=3, i=0^{\circ}$ and $z=3, i=45^{\circ}$ simulations. Here we show the results of the averaged $\sigma_*$ with stellar particles in the initial disk (the first and third columns), and without these particles (the second and fourth columns). All the lines and symbols are the same as in Fig.~\ref{fig:msigma_z2_d}.}
\label{fig:msigma_z3}
\end{figure*}

Figure~\ref{fig:msigma_z3} further proves this argument, from which we can see that results of the $z=3, i=0^{\circ}$ simulation are quite similar as shown in Figures~\ref{fig:msigma_z2_d} and \ref{fig:msigma_z2}. However, the resulting $\sigma_*$ have larger deviations from the $M_{\bullet}$-$\sigma_{*}$ relation after the dAGN stage, which may be caused by different inclined angles of the galaxy merger.

\section{Summary}
\label{sec:summary}

We perform three high resolution hydrodynamic simulations of galaxy mergers started at $z = 2$ and $z = 3$ to investigate the triggering of nuclear activities in merging galaxies and the associated evolution of relations between $M_{\bullet}$ and $M_{\mathrm{b}}$ or $\sigma_{*}$ during the merging processes.  

 We find that gas-rich major mergers can lead to the triggering of significant nuclear activities after the second and third pericentric passages. Significant accretion onto both MBHs can be triggered at the late stage of the mergers (after third or fourth pericentric passage). As a consequence of such nuclear active triggering, dAGNs appear on a separation of $\sim$\,kpc and typically last for a significant time period ($\sim 10 - 390$\,Myr).
 During the merging processes, the galactic bulge evolves with time due to the rapid star formation in each (or both) galactic center(s) and mixing of stars initially in galactic disks due to violent relaxation, and MBHs grow significantly due to accretion and finally merge to a more massive one. The growth of galactic bulge(s) and corresponding increases in its velocity dispersion(s) predate the growth of MBH(s) in the dAGN stages, and the MBHs in these stages deviate below the relation between MBH mass and bulge mass (or velocity dispersion), and they revert to the relations after the final mergers because significant accretion occurs mostly at separation less than a few kpc and the two MBHs merge with each other.    

\section*{Acknowledgements}
This work is supported by the National Key Program for Science and Technology Research and Development (grant no. 2016YFA0400704), the National Natural Science Foundation of China (grant no. 11690024 and 11873056), and the Strategic Priority Program of the Chinese Academy of Sciences (grant no. XDB 23040100).


\end{multicols}


\begin{thebibliography}{99}
\bibitem{1964ApJ...140..796S} E. E. Salpeter, ApJ, 140, 796 (1964)
%
\bibitem{1969Natur.223..690L} D. Lynden-Bell, Nature, 223, 690 (1969)
%
\bibitem{1999agnc.book.....K} J. H. Krolik,  Active galactic nuclei : from the central black hole to the galactic environment /Julian H. Krolik. Princeton, N. J. : Princeton University Press, c1999.,  (1999)
%
\bibitem{1998A&A...331L...1S} J. Silk, and  M. J. Rees, A\&A, 331, L1 (1998)
%
\bibitem{2003ApJ...596L..27K}  A. King, ApJL, 596, L27 (2003)
%
\bibitem{2012ARA&A..50..455F}  A. C. Fabian, ARA\&A, 50, 455 (2012)
%
\bibitem{2014ARA&A..52..589H}  T. M. Heckman, and  P.~N. Best, ARA\&A, 52, 589 (2014)
%
\bibitem{2015ARA&A..53..115K}  A. King, and  K. Pounds, ARA\&A, 53, 115 (2015)
%
\bibitem{1998AJ....115.2285M}  J. Magorrian,  S. Tremaine,  D. Richstone, R. Bender, G. Bower, A. Dressler, S. M. Faber, K. Gebhardt, R. Green, C. Grillmair, J. Kormendy, and T. Lauer,  AJ, 115, 2285 (1998)
%
\bibitem{2000ApJ...539L...9F}  L. Ferrarese, and D. Merritt,  ApJL, 539, L9 (2000)
%
\bibitem{2000ApJ...539L..13G} K. Gebhardt, R. Bender,  G. Bower, A. Dressler, S. M. Faber, A. V. Filippenko, R. Green, C. Grillmair, L. C. Ho, J. Kormendy, T. R. Lauer, J. Magorrian, J. Pinkney, D. Richstone, and S. Tremaine, ApJL, 539, L13 (2000)
%
\bibitem{2002ApJ...574..740T} S. Tremaine, K. Gebhardt, R. Bender, G. Bower, A. Dressler, S. M. Faber, A. V. Filippenko, R. Green, C. Grillmair, L. C. Ho, J. Kormendy, T. R. Lauer, J. Magorrian, J. Pinkney, and D. Richstone, ApJ, 574, 740 (2002)
%
\bibitem{2013ARA&A..51..511K} J. Kormendy, and  L. C. Ho, ARA\&A, 51, 511 (2013)
%
\bibitem{2016ASSL..418..263G} A. W. Graham, Galaxy Bulges and Their Massive Black Holes: A Review, edited by E. Laurikainen, R. Peletier and D. Gadotti, (Astrophysics and Space Science Library, 2016), pp. 263.
%
\bibitem{2005Natur.435..629S} V. Springel, S. D. M. White, A. Jenkins, C. S. Frenk, N. Yoshida, L. Gao, J. Navarro, R. Thacker, D. Croton, J. Helly, J. A. Peacock, S. Cole,  P. Thomas, H. Couchman, A. Evrard, J. Colberg, and F. Pearce,  Nature, 435, 629 (2005)
%
\bibitem{1989Natur.340..687H} L. Hernquist, Nature, 340, 687 (1989)
%
\bibitem{2011MNRAS.418.2043E} S. L. Ellison, D. R. Patton, J. T. Mendel and J.M. Scudder, MNRAS, 418, 2043 (2011)
%
\bibitem{2017ApJ...838..129B} R.S. Barrows, J.M. Comerford, J.E. Greene, and D. Pooley, ApJ, 838, 129 (2017)
%
 \bibitem{2018MNRAS.476.2308W} A. K. Weigel, K. Schawinski, E. Treister,  B.Trakhtenbrot, and D. B. Sanders, MNRAS, 476, 2308 (2018)
%
\bibitem{2012ApJ...758L..39T} E. Treister, K. Schawinski, C. M. Urry, and B. D. Simmons, ApJL, 758, L39 (2012)
%
\bibitem{2014A&A...569A..37M} N. Menci, M. Gatti, F. Fiore, and A. Lamastra, A\&A, 569, A37 (2014)
%
\bibitem{2014MNRAS.441.1297S} S. Satyapal, S. L. Ellison, W. McAlpine, R. C. Hickox, D. R. Patton, and J. T. Mendel, MNRAS, 441, 1297 (2014)
%
\bibitem{2015ApJ...804...34H} J. Hong, M. Im, M. Kim, and L. C. Ho, ApJ, 804, 34 (2015)
%
\bibitem{2018ApJ...853...63D} J. L. Donley, J. Kartaltepe, D. Kocevski, M. Salvato, P. Santini, H. Suh, F. Civano, A. M. Koekemoer, J. Trump, M. Brusa, C. Cardamone, A. Castro,  M. Cisternas, C. Conselice, D. Croton, N. Hathi, C. Liu, R. A. Lucas, P. Nair, D. Rosario, D. Sanders, B. Simmons, C. Villforth, D. M. Alexander, E. F. Bell, S. M. Faber, N. A. Grogin, J. Lotz, D. H. McIntosh, and T. Nagao, ApJ, 853, 63 (2018)
%
\bibitem{2018PASJ...70S..37G} A. D. Goulding, J. E. Greene, R. Bezanson, J. Greco, S. Johnson, A. Leauthaud, Y. Matsuoka, E. Medezinski, and A. M. Price-Whelan, PASJ, 70, S37 (2018)
%
\bibitem{2011ApJ...726...57C} M. Cisternas, K. Jahnke, K. J. Inskip, J. Kartaltepe, A. M. Koekemoer, T. Lisker, A. R. Robaina, M. Scodeggio, K. Sheth, J. R. Trump, R. Andrae, T. Miyaji, E. Lusso, M. Brusa, P. Capak, N. Cappelluti, F. Civano, O. Ilbert, C. D. Impey, A. Leauthaud, S. J. Lilly, M. Salvato, N. Z. Scoville, and Y. Taniguchi, ApJ, 726, 57 (2011)
%
\bibitem{2012ApJ...744..148K} D. D.Kocevski, S. M. Faber, M. Mozena, A. M. Koekemoer, K. Nandra, C. Rangel, E. S. Laird, M. Brusa, S. Wuyts, J. R. Trump, D. C. Koo, R. S. Somerville, E. F. Bell, J. M. Lotz, D. M. Alexander, F. Bournaud, C. J. Conselice, T. Dahlen, A. Dekel, J. L. Donley, J. S. Dunlop, A. Finoguenov, A. Georgakakis, M. Giavalisco, Y. Guo, N. A. Grogin, N. P. Hathi, S. Juneau, J. S. Kartaltepe, R. A. Lucas, E. J. McGrath, D. H. McIntosh, B. Mobasher, A. R. Robaina, D. Rosario, A. N. Straughn, A. van der Wel, and C. Villforth, ApJ, 744, 148 (2012)
%
\bibitem{2017MNRAS.470..755H} T. Hewlett, C. Villforth, V. Wild, J. Mendez-Abreu, M. Pawlik, and K. Rowlands, MNRAS, 470, 755 (2017)
%
\bibitem{2017MNRAS.465.2895L} E. K. Lofthouse, S. Kaviraj, C. J. Conselice, A. Mortlock, and W. Hartley, MNRAS, 465, 2895 (2017)
%
\bibitem{2019MNRAS.483.2441V}C. Villforth, H. Herbst, F. Hamann, T. Hamilton, C. Bertemes, A. Efthymiadou, and T. Hewlett, MNRAS, 483, 2441 (2019)
%
\bibitem{2003ApJ...582L..15K} S. Komossa, V. Burwitz, G. Hasinger, P. Predehl, J. S. Kaastra, and Y. Ikebe, ApJL, 582, L15 (2003)
%
\bibitem{2011ApJ...735L..42K} M. Koss, R. Mushotzky, E. Treister, S. Veilleux, R. Vasudevan, N. Miller, D. B. Sanders, K. Schawinski, and M. Trippe, ApJL, 735, L42 (2011)
%
\bibitem{2012ApJ...746L..22K} M. Koss, R. Mushotzky, E. Treister, S. Veilleux, R. Vasudevan, and M. Trippe, ApJL, 746, L22 (2012)
%
\bibitem{2012ApJ...748L...7V}  S. Van Wassenhove, M. Volonteri, L. Mayer, M. Dotti, J. Bellovary, and S. Callegari, ApJL, 748, L7 (2012)
%
\bibitem{2004ApJ...604L..33Z} H. Zhou, T. Wang, X. Zhang, X. Dong, and C. Li, ApJL, 604, L33 (2004)
%
\bibitem{2009ApJ...705L..76W} J. M. Wang, Y. M. Chen, C. Hu, W. M. Mao, S. Zhang, and W. H. Bian, ApJL, 705, L76 (2009)
%
\bibitem{2009ApJ...705L..20X} D. Xu, and S. Komossa, ApJL, 705, L20 (2009)
%
\bibitem{2009ApJ...698..956C} J. M. Comerford, B. F. Gerke, J. A. Newman, M. Davis, R. Yan, M. C. Cooper, S. M. Faber, D. C. Koo, A. L. Coil, D. J. Rosario, and A. A. Dutton, ApJ, 698, 956 (2009)
%
\bibitem{2010ApJ...708..427L} X. Liu, Y. Shen, M. A. Strauss, and J. E. Greene, ApJ, 708, 427 (2010)
%
\bibitem{2010ApJ...715L..30L} X. Liu, J. E. Greene, Y. Shen, and M. A. Strauss, ApJL, 715, L30 (2010)
%
\bibitem{2011ApJ...733..103F} H. Fu, A. D. Myers, S. G. Djorgovski, and L. Yan, ApJ, 733, 103 (2011)
%
\bibitem{2012ApJ...745...67F} H. Fu, L. Yan, A. D. Myers, A. Stockton, S. G. Djorgovski, G. Aldering, and J. A. Rich, ApJ, 745, 67 (2012)
%
\bibitem{2012ApJS..201...31G} J. Q. Ge, C. Hu, J. M. Wang, J. M. Bai, and S. Zhang, ApJS, 201, 31 (2012)
%
\bibitem{2013MNRAS.429.2594B} L. Blecha, A. Loeb, and R. Narayan, MNRAS, 429, 2594 (2013)
%
\bibitem{2016MNRAS.457.3878Z} X. G. Zhang, and L. L. Feng, MNRAS, 457, 3878 (2016)
%
\bibitem{2018ApJ...867...66C} J. M. Comerford, R. Nevin, A. Stemo, F. Müller-Sánchez, R. S. Barrows, M. C. Cooper, and J. A. Newman, ApJ, 867, 66 (2018)
%
\bibitem{2019MNRAS.482.1889W} M. X. Wang, A. L. Luo, Y. H. Song, S. Y. Shen, S. Feng, L. L. Wang, Y. F. Wang, Y. B. Li, B. Du, W. Hou,  Y. X.Guo, X. Kong, J. N. Zhang, MNRAS 482, 1889 (2019)
%
\bibitem{2019arXiv190406716W} M. X, Wang, and A. Luo, arXiv:1904.06716 (2019)
%
\bibitem{2011ApJ...737L..19C} J. M. Comerford, D. Pooley, B. F. Gerke, and G. M. Madejski, ApJL, 737, L19 (2011)
%
\bibitem{2011ApJ...740L..44F} H. Fu, Z. Y. Zhang, R. J. Assef, A. Stockton, A. D. Myers, L. Yan, S. G. Djorgovski, J. M. Wrobel, and D. A. Riechers, ApJL, 740, L44 (2011)
%
\bibitem{2012MNRAS.425.1185F} S. Frey, Z. Paragi, T. An, and K. {\'E}. Gab{\'a}nyi, MNRAS, 425, 1185 (2012)
%
\bibitem{2015ApJ...813..103M} F. M{\"u}ller-S{\'a}nchez, J. M. Comerford, R. Nevin, R. S. Barrows, M. C. Cooper, J. E. Greene, ApJ, 813, 103 (2015)
%
\bibitem{2005ApJ...627..721G} J. E. Greene, and L. C. Ho, ApJ, 627, 721 (2005)
%
\bibitem{2011ApJ...735...48S} Y. Shen, X. Liu, J. E. Greene, and M. A. Strauss, ApJ, 735, 48 (2011)
%
\bibitem{2017ApJ...848..126S} S. Satyapal, N. J. Secrest, C. Ricci, S. L. Ellison, B. Rothberg, L. Blecha, A. Constantin, M. Gliozzi, P. McNulty, and J. Ferguson, ApJ, 848, 126 (2017)
%
\bibitem{2018ApJ...854..169L} X. Liu, T. J. W. Lazio, Y. Shen, and M. A. Strauss, ApJ, 854, 169 (2018)
%
\bibitem{2018ApJ...862...29L} X. Liu, H. Guo, Y. Shen, J. E. Greene, and M. A. Strauss, ApJ, 862, 29 (2018)
%
\bibitem{2011ApJ...738...92Y} Q. Yu, Y. Lu, R. Mohayaee, and J. Colin, ApJ, 738, 92 (2011)
%
\bibitem{2005MNRAS.364.1105S} V. Springel, MNRAS, 364, 1105 (2005)
%
\bibitem{2003MNRAS.339..289S} V. Springel, and L. Hernquist, MNRAS, 339, 289 (2003)
%
\bibitem{2004MNRAS.348..435N} K. Nagamine, V. Springel, and L. Hernquist, MNRAS, 348, 435 (2004)
%
\bibitem{2014ApJ...780..145T} R. Thompson, K. Nagamine, J. Jaacks, and J. H. Choi, ApJ, 780, 145 (2014)
%
\bibitem{1939PCPS...35..405H} F. Hoyle, and R. A. Lyttleton, Proceedings of the Cambridge Philosophical Society, 35, 405 (1939)
%
\bibitem{1944MNRAS.104..273B} H. Bondi, and F. Hoyle, MNRAS, 104, 273 (1944)
%
\bibitem{1952MNRAS.112..195B} H. Bondi, MNRAS, 112, 195 (1952)
%
\bibitem{2009MNRAS.398...53B} C. M. Booth, and J. Schaye, MNRAS, 398, 53 (2009)
%
\bibitem{2009ApJ...690..802J} P.~H. Johansson, T. Naab, and A. Burkert, ApJ, 690, 802 (2009)
%
\bibitem{2014MNRAS.442.2751T} P. Taylor, and C. Kobayashi, MNRAS, 442, 2751 (2014)
%
\bibitem{2015MNRAS.446..521S} J. Schaye, R. A. Crain, R. G. Bower, M. Furlong, M. Schaller, T. V. C. Dalla, C. S. Frenk, I. G. McCarthy, J. C. Helly, A. Jenkins, Y. M. Rosas-Guevara, S, D. M. White, M. Baes, C. M. Camps, P. Navarro, J. F. Booth, Y. Qu, A. Rahmati, T. Sawala, P. A. Thomas, and J. Trayford, MNRAS, 446, 521 (2015)
%
\bibitem{2005MNRAS.361..776S} V. Springel, T. Di Matteo, and L. Hernquist, MNRAS, 361, 776 (2005)
%
\bibitem{2005Natur.433..604D} T. Di Matteo, V. Springel, and L. Hernquist, Nature, 433, 604 (2005)
%
\bibitem{2017SCPMA..60j9511Z}X. X. Zhang, and Y. J. Lu, Sci. China-Phys. Mech. Astron. 60, 109511 (2017), doi: 10.1007/s11433-017-9062-1
%
\bibitem{2005ApJ...630..705H} P. F. Hopkins, L. Hernquist, T. J. Cox,  T. Di Matteo, P. Martini, B. Robertson, and V. Springel, ApJ, 630, 705 (2005)
%
\bibitem{2017MNRAS.468.3395M} McAlpine, S., Bower, R. G., Harrison, C. M., R. A. Crain, M. Schaller, J. Schaye, and T. Theuns, MNRAS, 468, 3395 (2017)
%
\bibitem{2019MNRAS.483.4640W} E. X. Wang, P. Taylor, C. Federrath, and C Kobayashi, MNRAS, 483, 4640 (2019)
%
\bibitem{2005ApJ...622L...9S} V. Springel, and L. Hernquist, ApJL, 622, L9 (2005)
%
\bibitem{2017MNRAS.470.3946S} M. Sparre, and V. Springel, MNRAS, 470, 3946 (2017)
%
\bibitem{1990ApJ...356..359H} L. Hernquist, ApJ, 356, 359 (1990)
%
\bibitem{1998MNRAS.295..319M} H. J. Mo, S. Mao, and S. D. M. White, MNRAS, 295, 319 (1998)
%
\bibitem{2003ApJ...589L..21M} A. Marconi, and L. K. Hunt, ApJL, 589, L21 (2003)
%
\bibitem{2012ApJ...744....2A} A. Alonso-Herrero, M. Pereira-Santaella,  G. H.Rieke, and D. Rigopoulou, ApJ, 744, 2 (2012)
%
\bibitem{2019ApJ...870...31I} K. Ichikawa, C. Ricci, Y. Ueda, F. E. Bauer, T. Kawamuro, M. J. Koss, K. Oh, D. J. Rosario, T. T. Shimizu, M. Stalevski, L. Fuller, C. Packham, and B. Trakhtenbrot, ApJ, 870, 31 (2019)
%
\bibitem{2017MNRAS.469.4437C} P. R. Capelo, M. Dotti, M. Volonteri, L. Mayer, J. M. Bellovary, and S. Shen, MNRAS, 469, 4437 (2017)
%
\bibitem{1968adga.book.....S} J. L. Sersic, Atlas de Galaxias Australes (Observatorio Astronomico, Cordoba, Aggentina, 1968)
%
\bibitem{1970ApJ...160..811F} K. C. Freeman, ApJ, 160, 811 (1970)
%
\bibitem{2006ApJ...641...90R} B. Robertson, L. Hernquist, T. J. Cox, T. Di Matteo, P. F. Hopkins, P. Martini, and V. Springel, ApJ, 641, 90 (2006)
%
\bibitem{2006ApJ...641...21R} B. Robertson, T. J. Cox, L. Hernquist, M. Franx, P. F. Hopkins, P. Martini, and V. Springel, ApJ, 641, 21 (2006)
%
\bibitem{2012ApJ...747...33S} N. R. Stickley, and G. Canalizo, ApJ, 747, 33 (2012)
%
\bibitem{2007Sci...316.1874M} L. Mayer, S. Kazantzidis, P. Madau, M. Colpi, T. Quinn, and J. Wadsley, Science, 316, 1874 (2007)
%
\bibitem{2002MNRAS.331..935Y} Q. Yu, MNRAS, 331, 935 (2002)
%
\bibitem{2018ApJ...868...97K} F. M. Khan, P. R. Capelo, L. Mayer, and P. Berczik, ApJ, 868, 97 (2018)
%
\bibitem{2009ApJ...693.1554F}G. Foreman, M. Volonteri, and M. Dotti, ApJ. 693, 1554 (2009)
%
\bibitem{2013ApJ...777...64C} J. M. Comerford, K. Schluns, J. E. Greene, and R. J. Cool, ApJ, 777, 64 (2013)
%
\bibitem{2015ApJ...806..219C} J. M. Comerford, D. Pooley, R. S. Barrows, J. E. Greene, N. L. Zakamska, G. M. Madejski, M. C. Cooper, ApJ, 806, 219 (2015)
%
\bibitem{2018MNRAS.478.3056B} L. Blecha, G. F. Snyder, S. Satyapal, and S. L. Ellison, MNRAS, 478, 3056 (2018)
%
\end{thebibliography}
\end{document}